\documentclass[%
reprint,
superscriptaddress,
notitlepage,
 amsmath,amssymb,
 aps,
 prm,
]{revtex4-1}

\usepackage{graphicx}
\usepackage{dcolumn}
\usepackage{bm}
\usepackage{mhchem}

\begin{document}

\title{Deterministic strain-control of stability and current-induced motion of skyrmions in chiral magnets}

\author{Phuong-Vu Ong}
\affiliation{ Ames Laboratory, U.S. Department of Energy, Ames, Iowa 50011, USA }
\author{Tae-Hoon Kim}
\affiliation{ Ames Laboratory, U.S. Department of Energy, Ames, Iowa 50011, USA }
\author{Haijun Zhao}%
\affiliation{ Ames Laboratory, U.S. Department of Energy, Ames, Iowa 50011, USA }
\affiliation{ School of Physics, Southeast University, Nanjing 211189, China }
\author{Brandt A. Jensen}
\affiliation{ Ames Laboratory, U.S. Department of Energy, Ames, Iowa 50011, USA }
\author{Lin Zhou}
\email{linzhou@ameslab.gov}
\affiliation{ Ames Laboratory, U.S. Department of Energy, Ames, Iowa 50011, USA }%
\author{Liqin Ke}
\email{liqinke@ameslab.gov}
\affiliation{ Ames Laboratory, U.S. Department of Energy, Ames, Iowa 50011, USA }





\begin{abstract}
External magnetic field, temperature, and spin-polarized current are usually employed to create and control nanoscale vortex-like spin configurations such as magnetic skyrmions.
Although these methods have proven successful, they are not energy-efficient due to high power consumption and dissipation.
Coupling between magnetic properties and mechanical deformation, the magnetoelastic (MEL) effect, offers a novel approach to energy-efficient control of magnetism at the nanoscale.
It is of great interest in the context of ever-decreasing length scales of electronic and spintronic devices.
Therefore, it is desirable to establish a comprehensive framework capable of predicting effects of mechanical stress and enabling deterministic control of magnetic textures and skyrmions.
In this work, using an advanced scheme of multiscale simulations and Lorentz transmission electron microscopy measurements we demonstrate deterministic control of topological magnetic textures and skyrmion creation in thin films and racetracks of chiral magnets.
Our investigation considers not only uniaxial but also biaxial stress, which is ubiquitous in thin-film devices.
The biaxial stress, rather than the uniaxial one, was shown to be more efficient to create or annihilate skyrmions when the MEL coefficient and strain have the same or opposite signs, respectively.
It was also demonstrated to be a viable way to stabilize skyrmions and to control their current-induced motion in racetrack memory.
Our results open prospects for deployment of mechanical stress to create novel topological spin textures, including merons, and in control and optimization of skyrmion-based devices.

\end{abstract}

\maketitle


\section{Introduction}

Nanoscale vortex-like spin configurations, \textit{i.e.}, magnetic skyrmions\cite{Braun21},
are promising information carriers for future magnetic memories\cite{Kang16,Fert13}.
Investigations on these real-space topological states are usually devoted
to their creation or annihilation via external magnetic field,
temperature \cite{Muhlbauer09, Li13, Karube16, Tokunaga15, Peng18, Yu10, Guoqiang19, Anjan17},
and spin-polarized current control\cite{Jiang15}.
Although these methods have proven successful, they require high power consumption
with high power dissipation in the form of heat.
Coupling between magnetic properties and mechanical deformation, or magnetoelastic (MEL) effect,
promises energy-efficient control of magnetism at the nanoscale.
Moreover, the integration of skyrmions into current semiconductor technology
is highly likely to rely on thin-film heterostructures,
in which large stresses are often presented
due to the lattice mismatch between substrate and component layers.
Therefore, it is necessary to understand the effects of MEL coupling
and strain on magnetization dynamics, in general,
and on skyrmions, in particular, and the mechanism which ensures their stability in strained systems.

Studies have shown that uniaxial stress can deform skyrmions
from circular to elliptic shape in FeGe thin plates\cite{Shibata15}.
It can also fine-tune skyrmion crystals (SkXs) phase region
in MnSi\cite{Chacon15, Nii15} and Cu$_2$OSeO$_3$ crystals \cite{Seki17}.
Different mechanisms have been proposed to explain the observed phenomena,
including strained-induced magnetic anisotropy (MA) along the stress direction\cite{Butenko10, Nii15, JWang18}
and strained-induced anisotropy of Dzyaloshinskii–Moriya interaction(DMI) \cite{Shibata15}.
It is worth mentioning that even for the uniaxial stress,
the concomitant strain is induced in all three orthogonal directions.
As a result, an effective MEL field is induced not only in the direction
along the stress axis, but also perpendicular to it.
This is a unique feature of stress-induced MA,
and it is important to consider the combined effect of the effective MEL fields in all directions.
Moreover, these studies indicated that different types of stress, \textit{i.e.}, tensile or compressive
is required to stabilize skyrmions in different materials.
Therefore, it is crucial to establish a principle
that allows deterministic control of topological spin textures
in materials with given physical properties.
Lastly, biaxial stress, as opposed to the uniaxial one, is ubiquitous in modern electronic devices,
which are based on stacking of multilayers.
Therefore, a complete and practical understanding of the stress effect
should include the biaxial stress.

\begin{figure*}[t]
  \centering \includegraphics[scale=0.4]{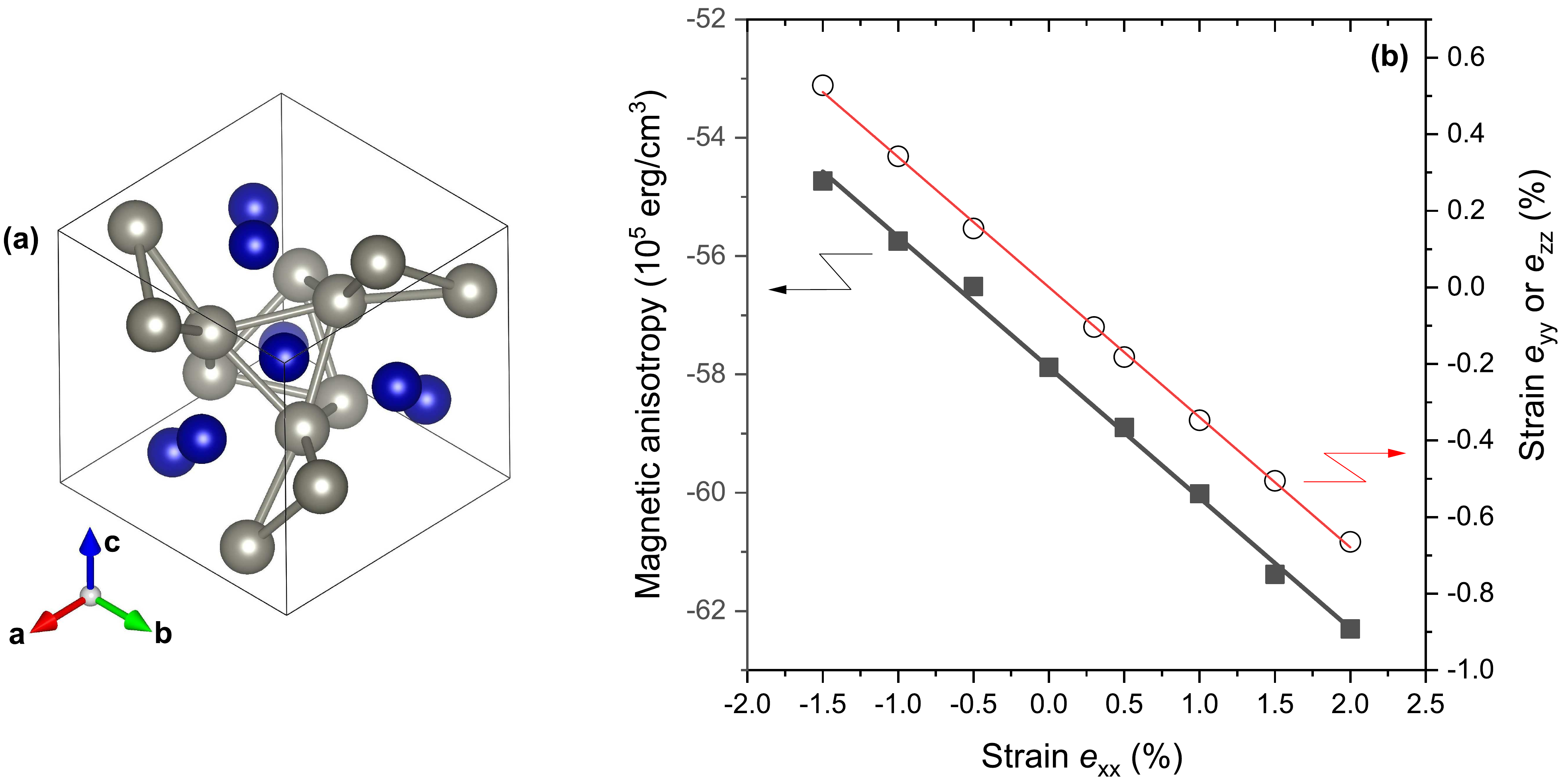}
  \caption{(a) Atomic structure of cubic Co$_{10}$Zn$_{10}$ viewed along a [111] direction.
    Blue balls indicate the 8$c$ sites, which are entirely occupied by Co atoms.
    Grey balls indicate the 12$d$ cites, which are occupied by Zn and the rest of Co atoms.
    (b) DFT results of magnetic anisotropy (left ordinate) and strain along the $y$ and $z$ directions
    ($e_{yy}$ and $e_{zz}$, respectively, on the right ordinate) as a function of strain along the $x$ direction ($e_{xx}$).}
  \label{DFTresult}
\end{figure*}

Here, we propose a theoretical framework for study of the stress effects,
which includes the second-order (in strain) MEL effect
and enables a deterministic control of topological spin textures and their dynamics.
Based on this framework, we report results
of advanced multiscale simulations
and Lorentz transmission electron microscopy observation
on the effects of both uniaxial and biaxial stress
on magnetization dynamics and skyrmions
in thin films and racetracks of chiral magnets.
MnSi and $\beta$-Mn-type Co-Zn-Mn were employed as two typical model materials.
Our results elucidate the relationship between MEL properties, types of stress, and
their effects. It was found that the biaxial stress, rather than the uniaxial one,
is more efficient to stabilize skyrmions and to control their dynamics in thin films and racetracks.
The results open prospects for deployment of mechanical stress to create
novel topological spin textures, including merons, and in control and optimization of
skyrmion-based devices.

\section{Theoretical and experimental details}

\subsection{Landau-Lifchitz-Gilbert equation with magnetoelastic interaction}

Suppose that a three dimensional magnetic sample occupies some domain
$\Omega\subset\mathbb{R}^3$ and let $\mathbf{m}$ be normalized magnetization
($\mathbf{m} = \mathbf{M}/|\mathbf{M}|$, where $|\mathbf{M}|=M_s$
is saturated magnetization). For systems with bulk Dzyaloshinskii–Moriya interaction(DMI),
the initial boundary value problem for the Landau-Lifchitz-Gilbert (LLG)
equation which we are interested takes the form

\begin{eqnarray}
\begin{cases}
  \label{eq:ibvp1}
  \partial_t\mathbf{m} - \alpha \mathbf{m} \times \partial_t\mathbf{m}
    =  - \mathbf{m} \times \mathbf{H}_{eff} \\
    \quad \text{in $\Omega\times(0,\infty)$,}\\
  \label{eq:ibvp2}
  2\ell_{ex}^2 \partial_{\mathbf{n}}\mathbf{m}
    = - \ell_{dm}\mathbf{m} \times \mathbf{n}  \quad \text{in $\Gamma\times(0,\infty)$,} \\
  \label{eq:ibvp3}
   \mathbf{m}(0)= \mathbf{m}^0  \quad \text{in $\Omega$.}
\end{cases}
\end{eqnarray}

\noindent where $\ell_{ex} =\sqrt{\frac{2A}{\mu_0M_s^2}}$ and  $\ell_{dm} = \frac{2D}{\mu_0M_s^2}$
are exchange and DMI lengths, respectively, in which $A$ and $D$ are
exchange interaction and DMI parameter, respectively;
$\alpha$ is the Gilbert damping parameter; $\mu_0$ is the vacuum magnetic permeability.
$\Gamma$ is the domain boundary and $\mathbf{m}^0$ is initial magnetization.
Note that the LLG equation in \eqref{eq:ibvp1} is in the dimensionless form.
The effective field, $\mathbf{H}_{eff}$, is determined by the first derivative of
total energy functional, $\mathcal{E}[\mathbf{M}]$,
with respect to the magnetization vector $\mathbf{M}$, more precisely

\begin{equation}\label{eq:Heff}
  \begin{split}
    \mathbf{H}_{eff} = & -\frac{1}{\mu_0M_s}\frac{\delta\mathcal{E}[\mathbf{M}]}{\delta\mathbf{M}}
      =  \ell_{ex}^2\triangle\mathbf{m} + \mathbf{H}_d + \mathbf{H}_{ext} \\
      & + Q(\mathbf{e}.\mathbf{m})\mathbf{e} - \ell_{dm}\mathbf{\nabla}\times\mathbf{m} + \mathbf{H}_{mel}
  \end{split}
\end{equation}
\noindent where $\mathbf{H}_d$ and $\mathbf{H}_{ext}$
are effective demagnetizing and external fields, respectively.
The descriptions of these conventional terms can be found
in Refs.~\cite*{Praetorius18, Alouges2014};
$Q=\frac{2K}{\mu_0M_s^2}$ is the MA coefficient,
in which $K$ is uniaxial MA energy.
We focus on the new term, $\mathbf{H}_{mel}$, which describes the MEL effect
and can be determined from the MEL energy functional\cite{Gilbert04, Eastman66}:

\begin{equation}\label{eq:melenergy}
  \begin{split}
    U_{mel}[\textbf{M}] = & \int\limits_{\Omega}\left( B_{1}\sum\limits_{i}m_i^2e_{ii}
      + \frac{1}{2}B_{2}\sum\limits_{i\neq j}m_im_je_{ij} \right. \\
      & \left. + \frac{1}{2}D_{11}\sum\limits_{i}m_i^2e_{ii}^2 \right)d\textbf{r}
  \end{split}
\end{equation}
\noindent where $m_i$, $i=$ $x$, $y$, or $z$, are the Cartesian component
of the normalized magnetization $\mathbf{m}$,
$B_1$ and $B_2$ are the first-order, and $D_{11}$ is the second-order MEL coefficients,
and $e_{ij}$ is strain tensor. Mechanical deformation induced by lattice mismatch usually
does not involve changes in crystallographic angles.
Therefore, shear strain will not be considered in the present work, \textit{i.e.}, $e_{ij}\approx 0$ for $i\neq j$.
With this in mind, we calculated the first functional derivative of the equation ~\eqref{eq:melenergy}
with respect to \textbf{M} and obtained the effective MEL field:

\begin{equation}\label{eq:melfield}
  \begin{split}
    \textbf{H}_{mel}[\textbf{M}]
    = & -\frac{1}{\mu_0M_s^2}\left [ ( 2B_1 + D_{11}e_{xx} ) m_x e_{xx}\hat{\textbf{x}} \right. \\
      & \left. + ( 2B_1 + D_{11}e_{yy} ) m_y e_{yy}\hat{\textbf{y}} \right. \\
      & \left. + ( 2B_1 + D_{11}e_{zz} ) m_z e_{zz}\hat{\textbf{z}} \right ]
  \end{split}
\end{equation}
\noindent where $\hat{\textbf{x}}$, $\hat{\textbf{y}}$, and $\hat{\textbf{z}}$
are the unit basis vectors of the Cartesian coordinate system.

For the FEA of MnSi skyrmion racetrack with the current-in-plane configuration,
the Zhang-Li spin torque term \cite{ZhangLi04} will
be added to the effective field \eqref{eq:Heff}:

\begin{eqnarray}
  \mathbf{H}_{ZL}[\textbf{M}] &=& [ \textbf{m} \times (\textbf{u}. \boldsymbol{\nabla}) \textbf{m} ] +
    \xi (\textbf{u}. \nabla) \textbf{m} \\
  \textbf{u} &=& \frac{1} {1+\xi^2} \frac{Pg_e\mu_B} {|e|\mu_0|\gamma| M_s^2} \textbf{j}_e
\end{eqnarray}
\noindent where $e$, $g_e$, and $\gamma$ are electron charge, $g-$factor,
and gyromagnetic ratio, respectively, $\mu_B$ is Bohr magneton.
Nonadiabatic spin torque parameter, $\xi$, is typically an order of
magnitude larger than the damping parameter $\alpha$ \cite{Garate09}.

\subsection{Uniaxial and biaxial strain in isotropic materials}

For an isotropic materials, we have the following relations between stress tensor ($e_{ij}$)
and strain tensor ($\sigma_{ij}$)\cite{Nye04},

\begin{eqnarray}
  \label{eq:stressstrainrel1}
  e_{xx} &=& \frac{1}{E}\left[ \sigma_{xx}-\nu(\sigma_{yy} + \sigma_{zz}) \right] \\
  \label{eq:stressstrainrel2}
  e_{yy} &=& \frac{1}{E}\left[ \sigma_{yy}-\nu(\sigma_{zz} + \sigma_{xx}) \right] \\
  \label{eq:stressstrainrel3}
  e_{zz} &=& \frac{1}{E}\left[ \sigma_{zz}-\nu(\sigma_{xx} + \sigma_{yy}) \right]
\end{eqnarray}

\noindent where $E$ is Young's modulus, $\nu$ is Poisson's ratio,
$\sigma_{xx}$, $\sigma_{yy}$, $\sigma_{zz}$ are stress along, $x$, $y$, and $z$ directions.

In the case of \textit{uniaxial} stress, $\sigma_{xx}=\sigma$ and all other stress components vanish.
From \eqref{eq:stressstrainrel1}-\eqref{eq:stressstrainrel3}, we have

\begin{eqnarray}
  \label{eq:ustressstrainrel1}
  e_{xx} &=& \frac{\sigma}{E} \\
  \label{eq:ustressstrainrel2}
  e_{yy} &=& -\nu e_{xx} \\
  \label{eq:ustressstrainrel3}
  e_{zz} &=& -\nu e_{xx}
\end{eqnarray}

In the case of \textit{biaxial} stress, $\sigma_{xx}=\sigma_{yy}=\sigma$
and $\sigma_{zz}=0$, we have,

\begin{eqnarray}
  \label{eq:bistressstrainrel1}
  e_{xx} &=& (1-\nu)\frac{\sigma}{E} \\
  \label{eq:bistressstrainrel2}
  e_{yy} &=&  e_{xx} \\
  \label{eq:bistressstrainrel3}
  e_{zz} &=& -\frac{2\nu}{1-\nu} e_{xx}
\end{eqnarray}

For metals, typical value of Poisson's ratio is $\nu\approx1/3$\cite{Gercek07}.
Therefore, from \eqref{eq:bistressstrainrel1}-\eqref{eq:bistressstrainrel3}
we have $e_{xx} = e_{yy}\approx -e_{zz}$ for the biaxial stress.

For materials with nonlinear MEL property such as MnSi, we performed finite element analysis (FEA)
of the problem \eqref{eq:ibvp1} using the \textit{Commics} (COmputational MicroMagnetICS) code\cite{Pfeiler18},
modified to include the MEL field~\eqref{eq:melfield}.
We are not aware of any micromagnetic codes,
which are capable of calculating the nonlinear MEL effects. The FEA was performed
using the (almost) second-order tangent plane scheme.
Within this scheme, numerical solution of the LLG has been proved to converge
toward the weak solution within an error of (almost) second-order
in time-step size\cite{Alouges2014, Fratta17}. The FEA implementation is based
on the multiphysics finite element software \textit{Netgen/NGSolve}\cite{ngsolve}.

For MnSi thin films and racetracks, the FEA was performed
using Delaunay tetrahedralization with maximal global mesh-size of 6.0~nm was employed.
Time-step size was set to 0.1~ps.
Thickness of the thin films and racetracks was fixed at 18.0~nm.
Damping parameter was $\alpha$ = 0.02.
Material physical parameters are obtained from the experiments for a 17.6-nm thick MnSi thin film\cite{Karhu12}.
In particular, exchange stiffness $A$ = $0.76\times10^{-12}$~J/m;
bulk DMI coefficient $D_M$ = $0.34\times10^{-3}$~J/m$^2$;
saturation magnetization $M_s$ = $0.16\times10^{6}$~A/m;
uniaxial MA $K$ = $0.9\times10^{4}$~J/m$^3$;
the first order and second order MEL coefficients were
$B_1$ = $1.0\times10^{6}$~J/m$^3$ and $D_{11} = -7.8\times10^{7}$ J/m$^3$, respectively.
It was found that most of these parameters are independent of or weakly dependent on thickness, except to uniaxial MA\cite{Karhu12}.

For simulations of MnSi racetrack with an applied in-plane spin-polarized current, we set $\xi$ = 0.2.
Degree of spin-polarization $P$ was set to the experimental value of 0.1 \cite{Neubauer09}.
Electron current density $j_e$ = 5.0$\times$10$^{10}$ A/m$^2$ was employed.
We note that $\textbf{j}_e$ denotes direction of electron motion,
which is opposite to motion of positive charge.

On the other hand, the alloy Co$_8$Zn$_{8.5}$Mn$_{3.5}$
is a linear MEL material (as shown below).
Moreover, it hosts skyrmions with relatively large radii and thus requires a large simulation area ($\sim\mu m$).
Therefore, micromagnetic simulations were carried out using the finite-difference code \textit{mumax}$^3$ \cite{Vansteenkiste14}.
The simulations were performed
on a sample of area 2.5$\times$2.5~$\mu m^2$ and thickness 190~nm
with discretization cell size of 5$\times$5$\times$190~nm$^3$.
External magnetic field was $B_{ext}$ = 0.15~T
and damping parameter was $\alpha$ = 0.02.
Cubic MA constant was set to $K$ = $0.5\times10^{4}$~J/m$^3$.
Dulk DMI coefficient was set to $D_M$ = $0.7\times10^{-3}$~J/m$^2$,
which is larger than the experimental value
for the Co$_8$Zn$_{8}$Mn$_{4}$ bulk ($0.53\times10^{-3}$~J/m$^2$)\cite{Takagi17},
but found to give the simulated period of the helical stripes
in agreement with our experimental value for the Co$_8$Zn$_{8.5}$Mn$_3$ thin plate.
For other material physical parameters,
we adopted the experimental values for
Co$_8$Zn$_{8}$Mn$_{4}$ alloy.
In particular, exchange stiffness $A$ = $9.2\times10^{-12}$~J/m\cite{Takagi17}
and saturation magnetization $M_s$ = $0.35\times10^{6}$~A/m\cite{Bocarsly19}.

\subsection{Density functional theory (DFT) calculation of the MEL coefficient}

To calculate the MEL coefficients and Poisson's ratio $\nu$ of the Co-Zn-Mn alloy,
we performed DFT calculations based on the projector augmented-wave formalism
for electron-ion potential\cite{Bloch94}, as implemented in
the Vienna \textit{Ab initio} Simulation Package\cite{KRESSE199615}.
Exchange-correlation interaction was treated within the Perdew-Burke-Ernzerhoff
functional form of the generalized gradient approximation (GGA)\cite{Perdew96}.
The binary parent alloy Co$_{10}$Zn$_{10}$ has a simple cubic structure
(space group number 213), in which the 8$c$ sites are mainly occupied by Co atoms
and the 12$d$ sites are mainly occupied by Zn and also randomly occupied by the remaining Co atoms (Fig.~\ref{DFTresult}a)\cite{Xie13}.
In a derived alloy Co$_x$Zn$_y$Mn$_z$ alloys ($x+y+z=20$), Mn atoms mainly occupy the 12$d$ sites,
but also share the 8c sites with Co atoms \cite{Nakajima19, Hori07}.
To avoid complications due to the randomness of site occupancies in the Co$_x$Zn$_y$Mn$_z$ alloys,
we employed a simplified model based on the binary alloy Co$_{10}$Zn$_{10}$.
The model has 10 Co atoms (8 at the 8$c$ sites and 2 at two arbitrary 12$d$ sites) and 10 Zn atoms at the 12$d$ sites.

Uniaxial stress was applied along the $x$ axis. At each strain $e_{xx}$ $=\varepsilon_u$,
geometry optimization, magnetic, and atomic relaxation were performed
until the change in the total energy between two ionic relaxation steps is smaller than $10^{-6}$~eV.
Then, the corresponding strain $e_{yy}$ and $e_{zz}$ along the $y$ and $z$ directions, respectively,
can be determined as a function of $e_{xx}$ (Fig.~\ref{DFTresult}b, right ordinate).
The MA was calculated as the total-energy difference between the two magnetic states
in which the magnetization is aligned along the [100] or [100] directions, respectively.
The obtained data is shown in Fig.~\ref{DFTresult}b (left ordinate).

In the phenomenological theory, MA can be determined by the following expression:
\begin{equation}\label{eq:MAdiff}
  \text{MA} = f_M(e_{ij}, m_x=1) - f_M(e_{ij}, m_z=1)
\end{equation}

\noindent where the magnetic energy density functional $f_M$ is expressed by:

\begin{equation}\label{eq:magenergy}
  \begin{split}
    f_M(e_{ij}, m_i) = & K(1-m_x^2)+ B_1 \sum \limits_{i=x,y,z} e_{ii}m_i^2 \\
      & + B_2\sum\limits_{i\neq j}e_{ij}m_im_j
  \end{split}
\end{equation}

\noindent Since shear strain will not be considered, \textit{i.e.}, $e_{ij}=0$ for $i\neq j$.
Substituting \eqref{eq:magenergy} into \eqref{eq:MAdiff} we obtain:

\begin{equation}\label{eq:MAdiff}
  \text{MA} = -K + B_1(e_{xx}-e_{zz})
\end{equation}

\noindent For uniaxial stress, $e_{zz} = -\nu e_{xx}$. Therefore, with putting $e_{xx} = \varepsilon_u$ we obtain

\begin{equation}\label{eq:MAfit}
 \text{MA} = -K + B_1(1+\nu)\varepsilon_u
\end{equation}

By fitting the $e_{yy(zz)}$ data (Fig.~\ref{DFTresult}b, right ordinate)
with the linear relations \eqref{eq:ustressstrainrel2} and \eqref{eq:ustressstrainrel3},
$\nu$ was found to be 0.34, which agrees with the typical Poisson's ratio for metals ($\approx1/3$).
By fitting the calculated MA data (Fig.~\ref{DFTresult}b, left ordinate) to \eqref{eq:MAfit},
we obtained a magnetoelastic coefficient value $B_1$ = $-1.65\times10^{6}$~J/m$^3$,
which is negative, as opposed to the positive value for MnSi.
As shown below, this result will be confirmed by our experimental results.

\begin{figure}[t]
  \centering \includegraphics[scale=0.27]{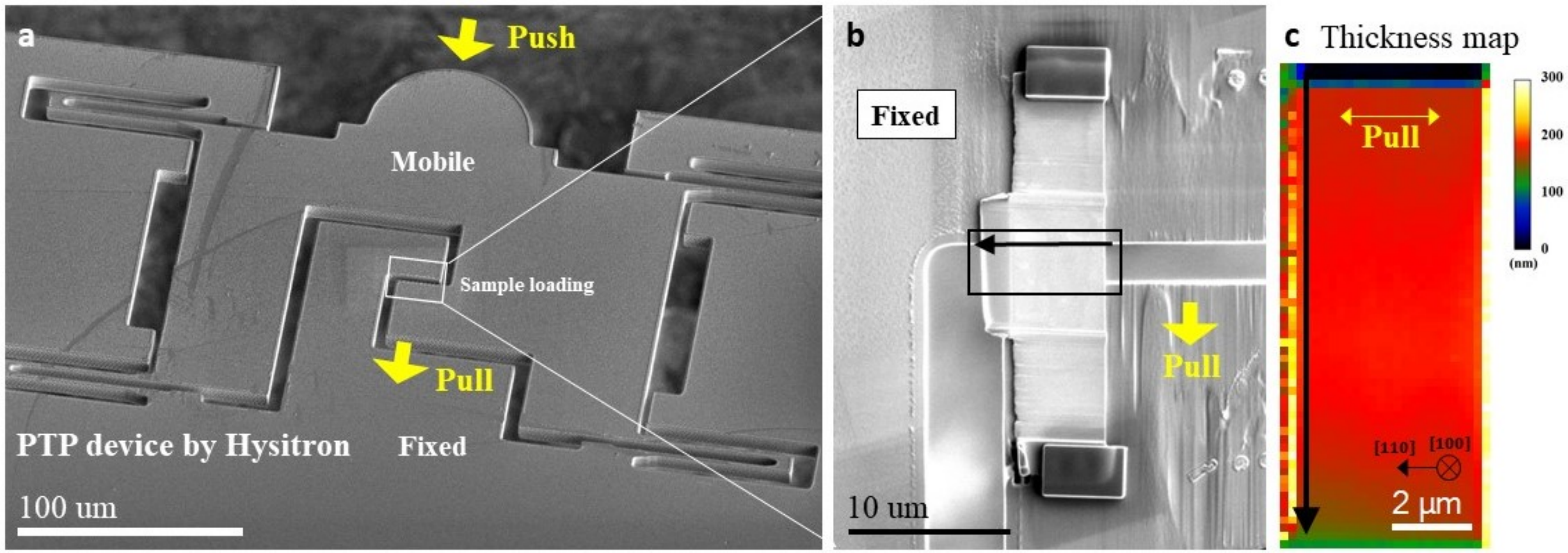}
  \caption{\textbf{TEM sample preparation for in-situ tensile testing.} (a) SEM image of push-to-pull (PTP) device.
    By pushing semi-circular part on top, tensile strain can be applied to the loaded thin plate.
    (b) (100)Co$_8$Zn$_{8.5}$Mn$_3$ thin plate loaded on the PTP device. Both ends of the plate were welded by carbon deposition.
    (c) Thickness map of the thin plate obtained by electron energy loss spectroscopy
    showing that the thin plate has a uniform thickness of $\sim$190 nm.}
  \label{exptfigS1}
\end{figure}

\subsection{Sample preparation and measurements}

\begin{figure*}[t]
  \centering
  \includegraphics[scale=0.5]{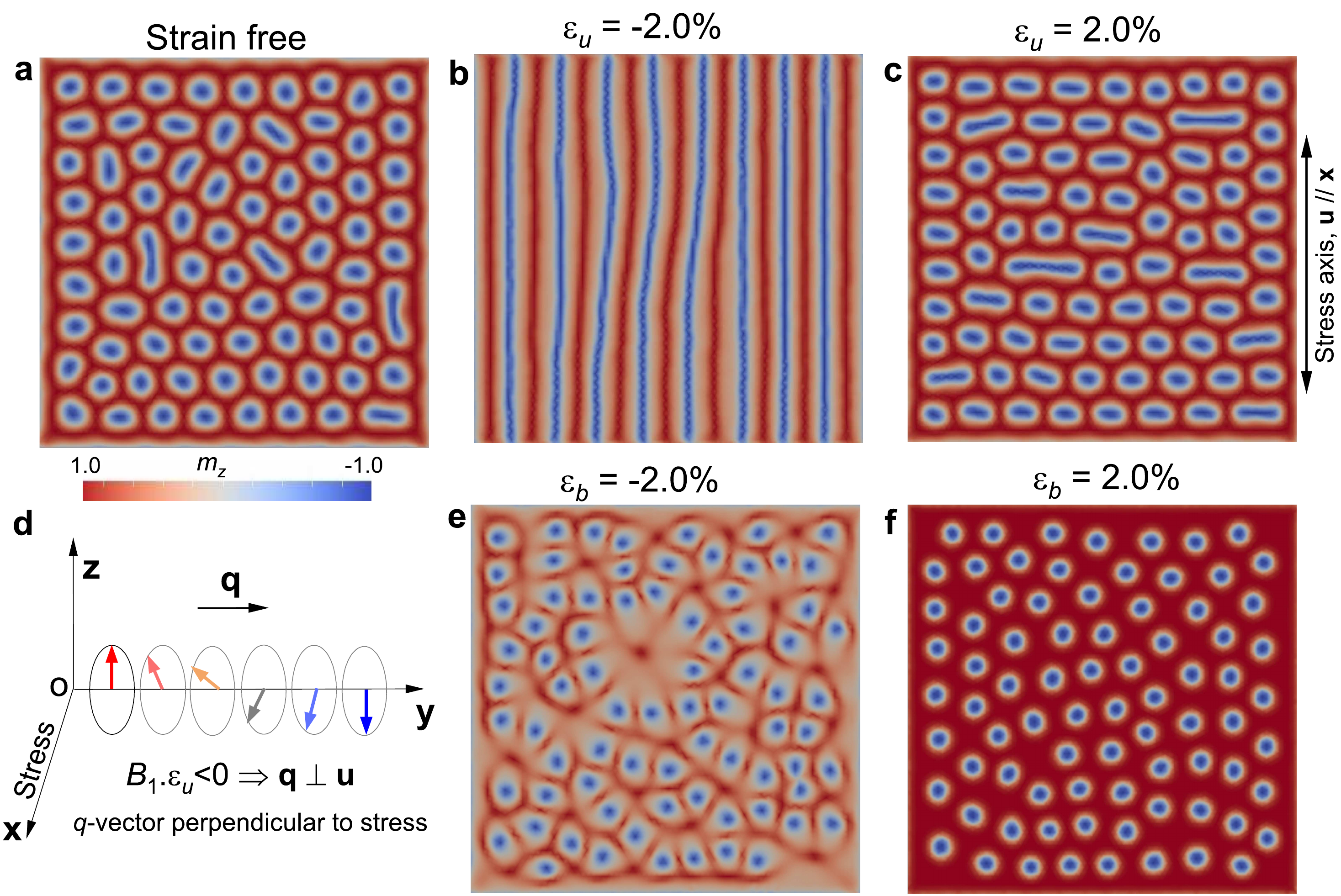}
  \caption{\textbf{Finite element analysis of the MnSi thin film under in-plane uniaxial and biaxial stress.}
  Sample area is 500$\times$500~nm$^2$ and thickness is 18~nm.
  External magnetic field is 0.2~mT along the film normal.
  (\textbf{a}, \textbf{b}, \textbf{c}) Equilibrium magnetization in the $xy$-plane
  viewed from the top under the uniaxial stress with $\varepsilon_u=$ 0, $-2.0$, and 2.0\%
  after the simulation duration of 15~ns, 5~ns, and 10~ns, respectively.
  The uniaxial stress axis (\textbf{u}) is along the $x$ axis, as indicated by the double headed arrow in (\textbf{c}).
  (\textbf{d}) The propagation vector $\textbf{q}$ of the helical phase is rendered perpendicular to the stress axis
  when the magnetoelastic coefficient $B_1$ and strain have opposite signs ($B_1\varepsilon_u<0$).
  (\textbf{e}) Magnetic texture featured by magnetization in the $xy$-plane in most of the perimeter region of the vortices,
  while still pointing downward in the core regions under the biaxial compressive stress ($\varepsilon_b=$ $-2$\%)
  after 10.0~ns.
  (\textbf{f}) Skyrmion crystal under the biaxial tensile stress ($\varepsilon_b=$ 2.0\%) after 35.3~ns.}
  \label{MnSifig}
\end{figure*}

The bulk Co$_8$Zn$_{8.5}$Mn$_{3.5}$ sample
was prepared by first sealing individual metals (all $>$ 99.9\% metals basis)
in a quartz ampoule backfilled with ultra-high purity argon.
The ampoule was placed into a furnace and heated to 1000 $^\circ$C for 12 hours,
then cooled at 1 $\circ$C/hr to 925 $^\circ$C and held for 96 hours before quenching into water.
Magnetic measurements were performed using a Quantum Design VersaLabTM vibrating sample magnetometer.
A thin, polycrystalline piece was polished so the sample dimensions were greater than 5:1 aspect ratio.
The sample was field cooled at a rate of 2 K/min under an applied field of $H =$ 20 Oe.
The polycrystalline sample has a Curie temperature of 330~K
and magnetization of 0.2~$\mu_B$/f.u. under an applied field of $H =$ 20 Oe.
A Co$_8$Zn$_{8.5}$Mn$_{3.5}$ (100) plate with [110] lateral orientation was fabricated using
the focused ion beam system FEI Helios NanoLab G3. The crystal orientation was examined
by electron backscattered diffraction analysis before lift-out.
The plate was thinned to approximately 190 nm thick and transferred on a push-to-pull (PTP)
device with 150 N/m$^2$ stiffness. Carbon deposition was conducted to ensure that the plate
was clearly adhered to the PTP device. To prevent Ga-ion-induced beam damage during the transfer,
low-kV Ga ion imaging was performed. Thickness map of the sample (Fig.~\ref{exptfigS1}c)
was obtained using electron energy loss spectroscopy with Gatan Quantum ER 965.

In-situ Lorentz transmission electron microscopy (LTEM) observation was
carried out on an FEI Tecnai G2-F20 operating at 200 kV of accelerating voltage.
A Hysitron PI 95 TEM PicoIndenter enables quantitative uniaxial tensile testing in the LTEM experiments.
Real-time applied force and displacement were measured.
By subtracting the PTP device portion from the measured force, actual force applied
to the specimen was calculated. As a result, a linear stress-strain relation
with Young's modulus of 85.52 GPa was obtained.
An external magnetic field was applied along the electron beam direction by partially
exciting the objective lens. The in-plane magnetization maps of magnetic
structures were obtained by the LTEM Fresnel images with a phase-retrieval
QPt software on the basis of the transport of intensity equation\cite{ishizuka_allman_2005}.

\section{Results and discussion}

\subsection{The uniaxial and biaxial stress effects on a nonlinear MEL chiral magnet}

Due to the lack of space-inversion symmetry, DMI is induced in MnSi
and makes it a material of choice for study of topological spin textures~\cite{Muhlbauer09}.
In a thin film form, MnSi exhibits nonlinear MEL behavior
with the first- and second MEL coefficients of $B_1$ = $1.0\times10^{6}$ J/m$^3$
and $D_{11} = -7.8\times10^{7}$ J/m$^3$, respectively\cite{Karhu12}.

Fig.~\ref{MnSifig} shows the FEA results for a 18-nm-thick MnSi thin film
under an external magnetic field of 0.2~T.
In the strain free condition, the system shows short stripes and skyrmions of various shapes
with an average diameter of about 58~nm (Fig.~\ref{MnSifig}a).

For a uniaxial stress along the $x$ axis,
the relationships between strains along the $x$, $y$, and $z$ axes
are $e_{xx}$ = $\varepsilon_u$ and $e_{yy}$ = $e_{zz}$ = $-\nu\varepsilon_u$,
as inferred from equations \eqref{eq:ustressstrainrel1}-\eqref{eq:ustressstrainrel3}.
Under the uniaxial \emph{compressive} stress ($\varepsilon_u$ = $-2.0$\%), these skyrmions elongate
and merge to form a helical phase with the spin-spiral propagation (or $q$-) vector
perpendicular to the stress direction (Fig.~\ref{MnSifig}b).
In contrast, the uniaxial \emph{tensile} stress does not
induce a helical phase, but only the elongation and alignment perpendicular to the stress axis (Fig.~\ref{MnSifig}c).

Underlying mechanism of the stress-induced formation of the helical phase
can be explained on the basis of strain-induced MA.
For a uniaxial stress along the $x$ axis, the effective MEL field \eqref{eq:melfield}
is equivalent to the following field up to addition by a field parallel to \textbf{m}

\begin{equation}\label{eq:melfielduni}
  \begin{split}
    \textbf{H}_{mel}^u[\textbf{M}]
      = & -\frac{1}{\mu_0M_s^2}\left [  2 (1+\nu) B_1\varepsilon_u \right. \\
        & \left. + (1-\nu^2) D_{11}\varepsilon_u^2 \right ] m_x \hat{\textbf{x}}
  \end{split}
\end{equation}

This effective MEL field has the same nature as a MA field along the $x$ axis.
We note that the field~\eqref{eq:melfielduni} can be made parallel to the stress direction ($x$)
only by combination of the effective MEL fields
induced not only in the direction along the stress axis, but also perpendicular to it.
Under the compressive stress, due to $B_1\varepsilon_u<0$ and $D_{11}<0$
the field~\eqref{eq:melfielduni} is parallel to $\hat{\textbf{x}}$
and thus induces an increase in the magnetization component $|m_x|$.
As a result, the skyrmions and short stripes are elongated along the stress direction.
These elongated skyrmions and stripes combine to form longer ones
and eventually helices (Fig.~\ref{MnSifig}b).
Due to the finite size along the thin film normal
a helical phase with $q$-vector oriented along the $z$ axis is prohibited.
The reason is that it will result in a lamellar structure of ferromagnetic layers
parallel to the $xy$-plane, which is a relatively unstable configuration.
As a result, the $q$-vector can be oriented only along the $y$ axis (Fig.~\ref{MnSifig}d).

\begin{figure}[t]
  \centering
  \includegraphics[scale=0.4]{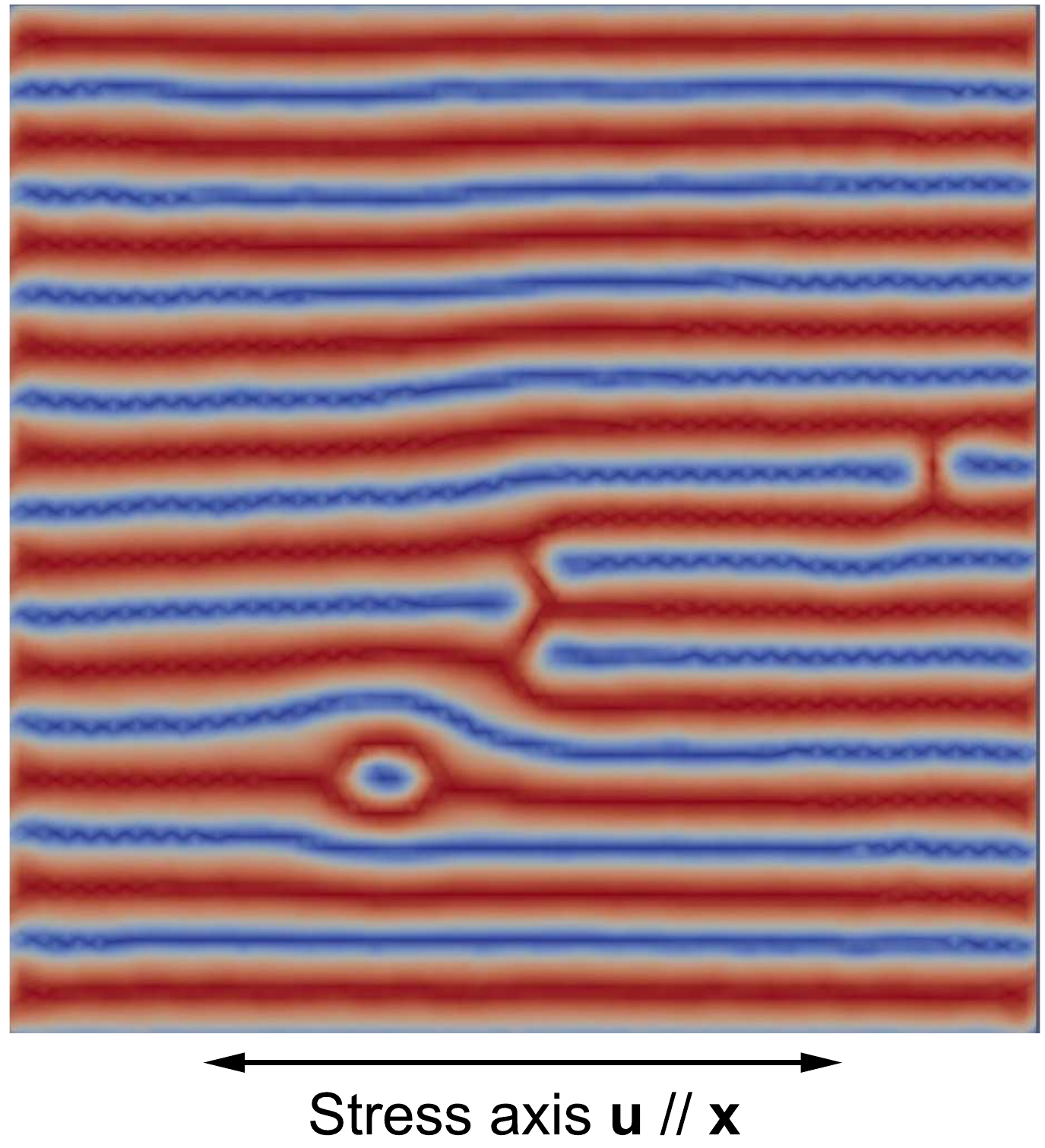}
  \caption{\textbf{Finite element analysis of the MnSi thin film under in-plane uniaxial stress.}
  Equilibrium magnetization in the $xy$-plane viewed from the top
  under the compressive uniaxial stress ($\varepsilon_u=$ $-2.0$\%).
  Sample area is 500$\times$500~nm$^2$ and thickness is 18~nm.
  External magnetic field is 200~mT along the film normal.
  Uniaxial stress axis (\textbf{u}) is along the \textbf{x} axis, as indicated by the double headed arrow.}
  \label{MnSifigD11zero}
\end{figure}

We found that including the second order term (due to the finite $D_{11}$)
is necessary for the formation of purely helical phase,
which otherwise contains discontinuous helices and an isolated skyrmion (Fig.~\ref{MnSifigD11zero}).
This second order effect can be understood in term of the additional field
$\left ( -\frac{(1-\nu^2)D_{11}\varepsilon_u^2}{\mu_0M_s^2} \right )m_x\hat{\textbf{x}}$ in~\eqref{eq:melfielduni},
which, for $D_{11}<0$, is parallel to $\hat{\textbf{x}}$ and thus enhances the in-plane MA.
Therefore, it further promotes the elongation and merging of skyrmions and hence the formation of helices.

On the other hand, for tensile stress ($B_1\varepsilon_u>0$)
the field~\eqref{eq:melfielduni} points opposite to $\hat{\textbf{x}}$.
Therefore, it drives magnetization equally likely in the $y$ and $z$ directions.
Due to the above mentioned asymmetry between the $y$ and $z$ axes,
skyrmion elongation occurs along the $y$ axis.
In this case, a helical phase could, however, not be formed (Fig.~\ref{MnSifig}c).
The reason is that the field~\eqref{eq:melfielduni}
now induces an enhancement in the perpendicular MA (PMA),
which tends to stabilize magnetization along the $z$ direction.
This is shown by the noticeable increase in the $m_z$ components,
as indicated by the enhanced color contrast in Fig.~\ref{MnSifig}c,
compared with that in Fig.~\ref{MnSifig}b.
Therefore, the enhanced PMA limits the elongation and suppresses the merging,
and thus stabilizes the (elongated) skyrmions .

For thin films, it is often more relevant to consider biaxial stress,
which is caused by the lattice mismatch between a thin film and the underlying substrate
or between the component layers in a heterostructure.
Under the biaxial stress, the $x$ and $y$ directions are equivalent
and, therefore, skyrmion elongation effect will not occur.
The effective MEL field \eqref{eq:melfield} is now equivalent to

\begin{equation}\label{eq:melfieldbiaxial}
  \begin{split}
    \textbf{H}_{mel}^b[\textbf{M}]
      = & \frac{1}{\mu_0M_s^2}\left [  2 \left ( \frac{1+\nu}{1-\nu} \right ) B_1\varepsilon_b \right. \\
        & \left. + \frac{ ( 1 + \nu ) ( 1 - 3 \nu ) } { (1-\nu)^2} D_{11}\varepsilon_b^2 \right ] m_z \hat{\textbf{z}}
  \end{split}
\end{equation}

\noindent where $\varepsilon_{b}$ is strain along the $x$ and $y$ directions
(see equations \eqref{eq:bistressstrainrel1}-\eqref{eq:bistressstrainrel3}).
Given that for most metals the Poisson's ratio $\nu$ is about $\approx1/3$,
the equation~\eqref{eq:melfieldbiaxial} implies that
the second order term is negligibly small, owing to the factor $( 1 - 3 \nu )$
and also the second order in strain ($\varepsilon_b^2$).
For the compressive stress ($B_1\varepsilon_b<0$), this field reduces PMA of the MnSi film.
As a result, magnetization is driven to the $xy$ plane in most of the perimeter regions of
the vortices, while still pointing downward in the core regions (Fig.~\ref{MnSifig}e).
This magnetic configuration resembles that of merons,
which are vortex-like spin textures with winding number $n=-1/2$\cite{Yu2018},
except that there are still finite out-of-plane magnetization components in some finite parts of the perimeters.
Nevertheless, the result gives rise to a possibility of creating merons
and their crystal with a biaxial stress if magnetic field and temperature can be fine-tuned into a proper condition.

Under the biaxial tensile stress ($B_1\varepsilon_b>0$), PMA is enhanced,
leading to a transformation of the short stripes into skyrmions
and a reduction of skyrmion diameter. After a simulation time of 10.0~ns,
the system was found to become all skyrmions with a diameter of about 50~nm.
These skyrmions were found to keep moving and reorganizing themselves in a relatively slow process.
Fig.~\ref{MnSifig}f shows the skyrmion system after 35.3~ns.
It is expected that a hexagonal SkX will form after a sufficiently long simulation time.

\begin{figure*}[t]
  \centering
  \includegraphics[scale=0.6]{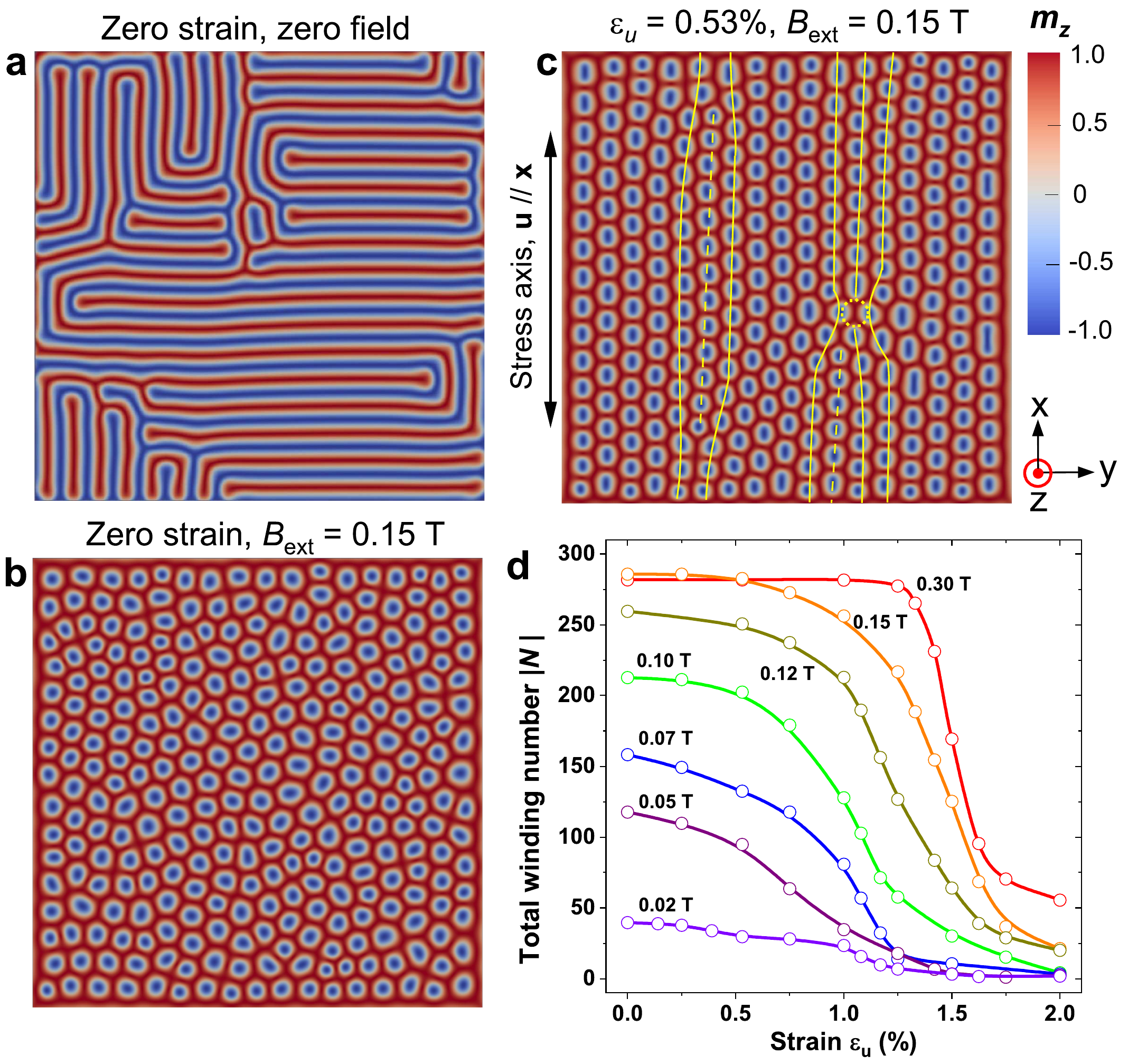}
  \caption{\textbf{Finite difference simulations of the Co$_8$Zn$_{8.5}$Mn$_{3.5}$
  thin plate under in-plane uniaxial stress.}
  Sample area and thickness are 2.5$\times$2.5~$\mu m^2$ and 190~nm, respectively. Simulation duration is 100ns.
  (\textbf{a}, \textbf{b}) Helical and SkX phases in the $xy$-plane
  viewed from the top in the zero strain condition
  under the external magnetic field $B_{ext}=$ 0 and 0.15~T, respectively, along the $z$ direction.
  (\textbf{c}) Elongation and alignment of skyrmions
  at strain $\varepsilon_u=$ 0.53\% under the field $B_{ext}=$ 0.15~T.
  The uniaxial stress axis (\textbf{u}) is along the $x$ axis, as indicated by the double headed arrow.
  Solid lines indicate lattice distortion due to edge dislocations (dashed lines)
  or skyrmion vacancy (dotted circle).
  (\textbf{d}) Total winding number of the system as a function of strain $\varepsilon_u$ and magnetic field.}
  \label{CoZnMnfig}
\end{figure*}

\begin{figure*}[t]
  \centering
  \includegraphics[scale=0.6]{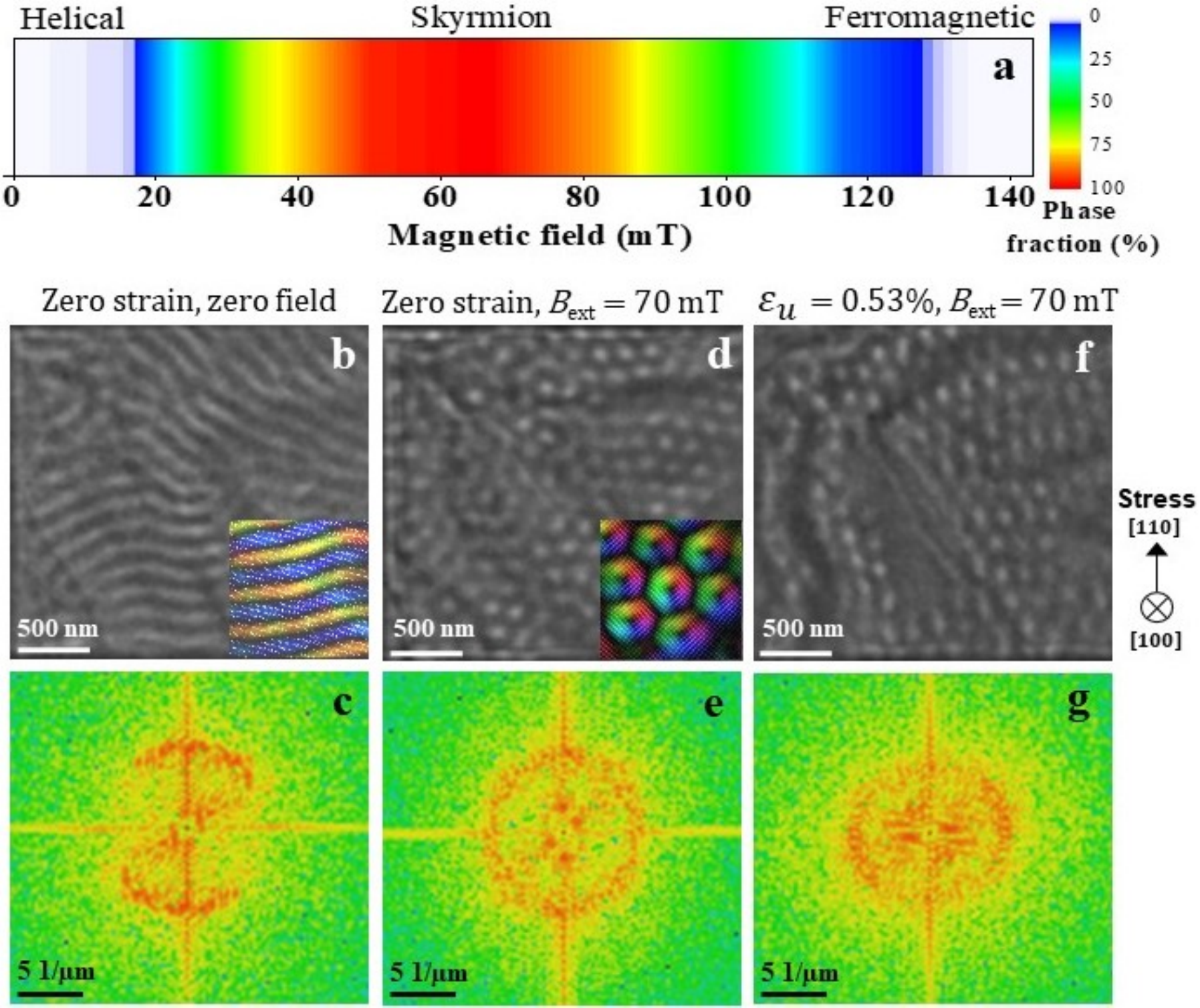}
  \caption{\textbf{Phase diagram and magnetic textures in the 190 nm (100) Co$_8$Zn$_{8.5}$Mn$_3$
  thin plate and effect of uniaxial stress along a [110] direction.}
  (\textbf{a}) A contour plot of skyrmion density at room temperature under the strain free condition
  as deduced from LTEM observation. Color scale indicates skyrmion phase fraction (in \%) of the observed area.
  (\textbf{b,c}) Under-focused LTEM image of helical spin texture
  and the corresponding fast Fourier transform (FFT), respectively, in a strain free sample under zero field.
  (\textbf{d,e}) Skyrmion crystal and its FFT, respectively, in a strain free sample under 70~mT.
  (\textbf{f}) Skyrmions elongated and aligned along the uniaxial stress direction [110] under 70~mT
  and strain $\varepsilon_u =$ 0.53\%.
  (\textbf{g}) The corresponding FFT of (\textbf{f}) exhibits an elliptic shape
  with major-axis perpendicular to the stress direction.}
  \label{exptfig}
\end{figure*}

\begin{figure*}[t]
  \centering
  \includegraphics[scale=0.44]{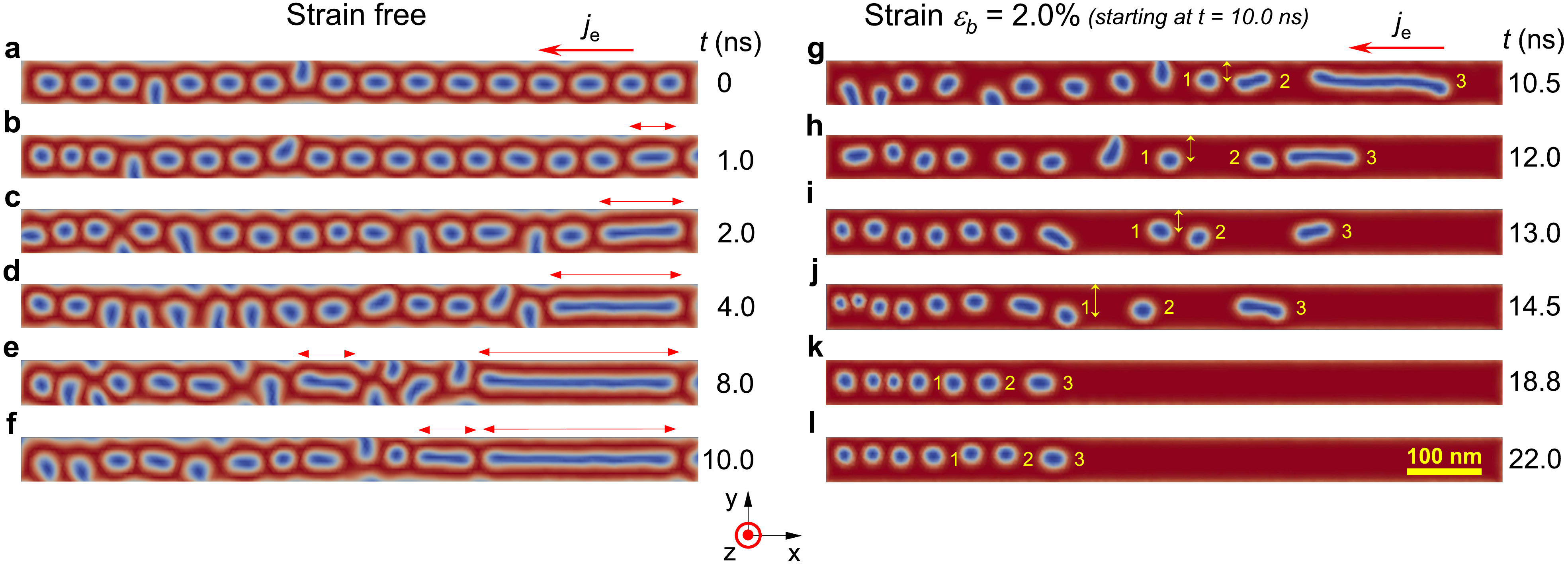}
  \caption{\textbf{Finite element analysis of magnetization dynamics on MnSi racetrack.}
  The racetrack is 900-nm long, 60-nm wide, and 18-nm thick.
  External perpendicular magnetic field is 0.15~T.
  In-plane electron current density is $j_e$ = 5.0$\times10^{11}$ A/m$^2$, passed from right to left.
  (\textbf{a-f}) Snapshots of magnetization dynamics in a strain free racetrack. Double headed arrows
  indicate current-induced elongation of skyrmions and magnetic stripe.
  Corresponding simulation time ($t$) is shown at the right end of each panel.
  (\textbf{g-l}) Shrinking of the stripe to a skyrmion and current-induced motion of skyrmions
  when a biaxial stress ($\varepsilon_{b} = -2.0\%$) is applied at $t=10$~ns.
  Numerals indicate the stripe or skyrmions which have longest free paths.
  Vertical arrows indicate transverse motion of the skyrmion 1 to the lower edge of the racetrack.}
  \label{MnSiRacetrack}
\end{figure*}

As discussed above, the out-of-plane (along the $z$ axis) MEL field
plays a key role in stabilization, creation, or annihilation of skyrmions.
Comparing the first order terms in the equations~\eqref{eq:melfielduni}
and ~\eqref{eq:melfieldbiaxial} for the uniaxial and biaxial stress, respectively,
reveals a difference by a factor of $\left ( \frac{1}{1-\nu} \right )$.
Most of materials have $\nu$ ranging between 0.0 and 0.5.
For the typical case of $\nu = 1/3$, this translates into a strain-induced PMA either larger or smaller
by 50\% for the biaxial stress, compared with that induced by the corresponding uniaxial one.
Based on the above discussed results, we proposed the following criteria for efficient creation
and control of skyrmions with mechanical stress: \textit{i}) the applied stress should be biaxial;
\textit{ii}) the MEL coefficient $B_1$ and strain have to be in the same sign.
Moreover, due to the factor $\left ( \frac{1}{1-\nu} \right )$
the relative efficiency of the biaxial stress depends on the Poisson's ratio $\nu$.
A higher $\nu$ value leads to a higher efficiency of the mechanical control
with biaxial stress, as opposed to the uniaxial one.

\subsection{The magnetoelastic sign effect on skyrmion elongation and helical phase formation}

As can be inferred from the effective MEL fields~\eqref{eq:melfielduni} and \eqref{eq:melfieldbiaxial},
the strain $\varepsilon_{u(b)}$ always appears in the forms of products
$B_1\varepsilon_{u(b)}$ and $D_{11}\varepsilon_{u(b)}^2$.
From the above mentioned mechanism of stress effect,
it is clear that formation and orientation of the elongated skyrmions
and helical phase relative to the stress direction
depend on sign of the MEL coefficient $B_1$. Therefore, they can vary from one material to another.

To further demonstrate the deterministic control of magnetic texture,
we apply our micromagnetic framework to a $\beta$-Mn-type Co$_x$Zn$_y$Mn$_z$ alloy with $x+y+z=20$.
This class of chiral magnets can host skyrmions beyond room temperature~\cite{Tokunaga15}.
As shown above,  our DFT calculations predict that the $\beta$-Mn-type Co-Zn-Mn alloy
has linear MEL behavior with the negative MEL coefficient $B_1$,
\textit{i.e.}, opposite to that for MnSi,
and thus makes it an ideal choice for a further test of the framework.

To obtain the magnetic texture in the zero magnetic field
and strain free condition, we carried out simulation
at high temperature and then slowly decreased it to zero,
\textit{i.e.}, employed simulated annealing simulation
to accelerate the convergence process toward the ground state.
The magnetic texture exhibits discontinuous helical stripes with different orientations
and an average stripe period of about 156~nm (Fig.~\ref{CoZnMnfig}a).

Under the external magnetic field $B_{ext}=$ 0.15~T along the thin plate normal,
polydomain SkX is formed with an average skyrmion diameter of about 150~nm (Fig.~\ref{CoZnMnfig}b).
A uniaxial tensile stress with $\varepsilon_u=$ 0.53\% is then applied along the $x$ direction.
The stress effect is twofold: first, it induces elongation of skyrmions;
and second, it renders skyrmions aligned along the stress directions,
and thus induces a polydomain-to-monodomain phase transition (Fig.~\ref{CoZnMnfig}c).
The monodomain SkX has edge dislocations (dashed lines)
and a skyrmion vacancy (dotted circle).
It is worth mentioning that the local distortion
induced by the vacancy is especially strong for the SkX,
in sharp contrast to that for a solid crystal.

A topological magnetic texture is characterized
by a winding number or topological charge
defined as

\begin{equation}
  n=\frac{1}{4\pi}\int \mathbf{m}.(\partial_x \mathbf{m}
    \times \partial_y \mathbf{m})dxdy
\end{equation}

\noindent which is a negative integer for the Co-Zn-Mn thin plate
under the external magnetic field parallel to the $z$ axis.
Fig.~\ref{CoZnMnfig}d shows absolute value of total winding number $|N|$
of the system as a function of strain and magnetic field below 0.3~T.
Three variation regimes of $|N|$ can be identified: first,
it is slowly decreased or unaltered with increasing strain;
then at a certain strain value depending on the applied field, it is rapidly reduced;
after that, it slowly decreases toward zero. In the first regime,
the main processes are elongation and alignment of skyrmions and short stripes,
which do not involve change in topology of the magnetic texture.
In the second regime, merging of elongated skyrmions and short helical stripes becomes active.
Moreover, in this regime strain is also sufficiently large to
align magnetization along the stress direction in most of the thin plate area,
leading to local transitions to ferromagnetic (FM) states, which are topologically trivial.
These effects together cause the rapid decrease in $|N|$.
For strong magnetic fields on the verge of a FM transition ($\sim$0.30~T),
the stress-induced elongation and merging of skyrmions are suppressed
for strain below a critical value.
Above the critical strain, the system changes directly from a skyrmion phase
to a FM phase. Therefore, for the magnetic field of 0.3~T
the total winding number $|N|$ falls sharply at the strain value $\varepsilon_u=1.38$\%.

For the Co-Zn-Mn thin plate, the elongation and alignment of skyrmions (Fig.~\ref{CoZnMnfig}c)
is at 90 degree with respect to those for the MnSi thin film under the same type of strain (tensile, Fig.~\ref{MnSifig}c).
This is due to the opposite signs of the first order MEL coefficients for the two materials.
Based on the results for both of the systems, the following rule can be drawn:
for a thin plate or thin film under uniaxial stress,
if the product $B_1\varepsilon_u$ is negative, skyrmion elongation and alignment
will be parallel to the stress direction.
Moreover, the stress diminishes PMA of the system and thus tends to induce a helical phase
with $q$-vector perpendicular to the stress direction.
On the other hand, if $B_1\varepsilon_u$ is positive, the elongation
and alignment will be perpendicular to the stress direction.
Moreover, the elongation will be limited due to the stress-induced enhancement of PMA.

\subsection{Experimental confirmation}

Based on LTEM observation at room temperature, a phase diagram was established
for the observed rectangular area under increasing external magnetic field,
$B_{ext}$. Transitions from helical to skyrmion phase
and from skyrmion to FM phase occurs at $B_{ext}\sim$ 20 and 130~mT, respectively (Fig.~\ref{exptfig}a).
Fig.~\ref{exptfig}b shows helical texture in the zero field condition.
These helical stripes have different orientations and average period of about 160~nm,
resembling the simulation result for the zero field and zero strain condition (Fig.~\ref{CoZnMnfig}a).
Under $B_{ext}=$ 70~mT, a SkX can be observed
with average skyrmion diameter about 140~nm (Fig.~\ref{exptfig}d),
compared to the simulation result of 150~nm.

Elliptic skyrmions aligned along the stress direction can be observed
at strain $\varepsilon_u =$ 0.53\% (Fig.~\ref{exptfig}f).
The corresponding fast Fourier transform of the LTEM image
has an elliptic shape with the major axis perpendicular to the stress direction (Fig.~\ref{exptfig}g),
confirming the skyrmions are elongated along the stress direction.
This agreement between the experimental and simulation results also confirms
the DFT prediction of the negative MEL coefficient $B_1$ for this alloy.

\subsection{Stress control of skyrmion motion on MnSi racetrack}

To demonstrate a practical application of our findings,
we consider a 900-nm long, 60-nm wide, and 18-nm thick MnSi racetrack
under a perpendicular magnetic field of 0.15~T.
An in-plane electron-current $j_e=$5.0$\times10^{11}$~A/m$^2$ is passed through the track
along the long ($x$) axis. The current is found to induce elongation of a skyrmion.
The skyrmion can grow into a long stripe of up to 260~nm in 10.0~ns (Fig.\ref{MnSiRacetrack}b-f).
Shorter stripes are also induced around the middle of the racetrack (Fig.\ref{MnSiRacetrack}e and f).
As a result, the current-induced motion of skyrmions is suppressed.
As the stability of skyrmions and their steady flow are essential prerequisites
for operation of skyrmion racetrack devices, such skyrmion elongation and magnetic stripes would
have a detrimental effect on the devices.

Under the biaxial compressive stress ($\varepsilon_{b} = -2\%$), these stripes are
transformed into skyrmions (indicated by 2 and 3 in Fig.\ref{MnSiRacetrack}g-l).
As a result, the skyrmions are unblocked and start to flow following the electron current.
The skyrmions 1 and 2 are in free motion during time between $t=13.0$ to 14.5~ns (Fig.\ref{MnSiRacetrack}i and j)
and $t=12.0$ to 13.0~ns (Fig.\ref{MnSiRacetrack}h and i), respectively.
We note that their motion also has finite
transverse components (indicated by the double headed arrows for the skyrmion 1 in Fig.\ref{MnSiRacetrack}g-j).
This is the observed skyrmion Hall effect caused by the magnetic Magnus force,
which is induced by the skyrmion topological charge
and is perpendicular to the current and the magnetic field directions\cite{Jonietz1648, Jiang2016Magnus}.
The transverse motion is limited due to the narrow width and reflection at the racetrack boundary.
Average horizontal velocity is estimated to be 80.0~m/s,
which is comparable to the experimental and predicted values
for \ce{Pt}/\ce{CoFeB}/\ce{MgO} and Pt/Co thin films, respectively\cite{Woo16, Sampaio13}.

Stability of skyrmions is a major challenge for their applications at room-temperature.
Our results demonstrate that the biaxial stress can be employed to efficiently
create, stabilize, and control skyrmions.
This can be achieved, for example, by growth of the racetrack on a piezoelectric substrate,
which will facilitates an energy-efficient control of strain via an applied voltage or electric field.
This strain-mediated approach has already been proved to assist magnetization switching
in magnetoelectric memories \cite{Nan19, Hu15, Buzzi13}.
Such a device structure can be employed to control
and enhance stability of skyrmion-based racetrack memories and sensors.
Another advantage of such a multiferroic heterostructure is that skyrmions
can also be controlled by manipulating the ferroelectric polarization of the substrate.


To summarize, we have demonstrated theoretically
and experimentally the mechanical control of topological magnetic textures
and their dynamics in thin films and thin plates of chiral magnets.
Based on the magnetoelastic coupling, our theoretical framework
elucidates the relationship between magnetoelastic properties of materials,
types of stress (tensile or compressive), and their effects,
and thus enables a deterministic control of the topological spin textures.
It was found that the biaxial stress, rather than the uniaxial one,
is more efficient to annihilate, create, and stabilize skyrmions.
Moreover, creation or annihilation occurs when the magnetoelastic coefficient
and strain have the same or opposite signs, respectively.
Stress can induce a rich variety of topological spin textures,
possibly including merons.
Biaxial stress was also demonstrated to be a viable way to stabilize skyrmions
and to control their current-induced motion in racetrack memory.
We hope our findings stimulate further research and open prospects for deployment of mechanical stress
in control and optimization of skyrmion-based devices.

\begin{acknowledgments}
 \noindent This work is supported in part by Laboratory Directed Research and Development (LDRD) funds
 through Ames Laboratory (P.-V.O., T.K., H.Z., L.Z.) and by the U.S. Department of Energy,
 Office of Science, Basic Energy Sciences, Materials Science and Engineering Division.
 Ames Laboratory is operated for the U.S. Department of Energy by Iowa State University under
 Contract No. DE-AC02-07CH11358. L.K. was supported by the U.S. Department of Energy,
 Office of Science, Office of Basic Energy Sciences, Materials Sciences and Engineering Division,
 Early Career Research Program following conception and initial work supported by LDRD.
 All TEM and related work were performed using instruments in the Sensitive Instrument Facility in Ames Lab.
\end{acknowledgments}

\bibliography{strain_sk_ref_aps}

\begin{thebibliography}{49}%
\makeatletter
\providecommand \@ifxundefined [1]{%
 \@ifx{#1\undefined}
}%
\providecommand \@ifnum [1]{%
 \ifnum #1\expandafter \@firstoftwo
 \else \expandafter \@secondoftwo
 \fi
}%
\providecommand \@ifx [1]{%
 \ifx #1\expandafter \@firstoftwo
 \else \expandafter \@secondoftwo
 \fi
}%
\providecommand \natexlab [1]{#1}%
\providecommand \enquote  [1]{``#1''}%
\providecommand \bibnamefont  [1]{#1}%
\providecommand \bibfnamefont [1]{#1}%
\providecommand \citenamefont [1]{#1}%
\providecommand \href@noop [0]{\@secondoftwo}%
\providecommand \href [0]{\begingroup \@sanitize@url \@href}%
\providecommand \@href[1]{\@@startlink{#1}\@@href}%
\providecommand \@@href[1]{\endgroup#1\@@endlink}%
\providecommand \@sanitize@url [0]{\catcode `\\12\catcode `\$12\catcode
  `\&12\catcode `\#12\catcode `\^12\catcode `\_12\catcode `\%12\relax}%
\providecommand \@@startlink[1]{}%
\providecommand \@@endlink[0]{}%
\providecommand \url  [0]{\begingroup\@sanitize@url \@url }%
\providecommand \@url [1]{\endgroup\@href {#1}{\urlprefix }}%
\providecommand \urlprefix  [0]{URL }%
\providecommand \Eprint [0]{\href }%
\providecommand \doibase [0]{https://doi.org/}%
\providecommand \selectlanguage [0]{\@gobble}%
\providecommand \bibinfo  [0]{\@secondoftwo}%
\providecommand \bibfield  [0]{\@secondoftwo}%
\providecommand \translation [1]{[#1]}%
\providecommand \BibitemOpen [0]{}%
\providecommand \bibitemStop [0]{}%
\providecommand \bibitemNoStop [0]{.\EOS\space}%
\providecommand \EOS [0]{\spacefactor3000\relax}%
\providecommand \BibitemShut  [1]{\csname bibitem#1\endcsname}%
\let\auto@bib@innerbib\@empty
\bibitem [{\citenamefont {Braun}(2012)}]{Braun21}%
  \BibitemOpen
  \bibfield  {author} {\bibinfo {author} {\bibfnamefont {H.-B.}\ \bibnamefont
  {Braun}},\ }\bibfield  {title} {\bibinfo {title} {Topological effects in
  nanomagnetism: from superparamagnetism to chiral quantum solitons},\
  }\href@noop {} {\bibfield  {journal} {\bibinfo  {journal} {Adv. Phys.}\
  }\textbf {\bibinfo {volume} {61}},\ \bibinfo {pages} {1} (\bibinfo {year}
  {2012})}\BibitemShut {NoStop}%
\bibitem [{\citenamefont {Kang}\ \emph {et~al.}(2016)\citenamefont {Kang},
  \citenamefont {Huang}, \citenamefont {Zhang}, \citenamefont {Zhou},\ and\
  \citenamefont {Zhao}}]{Kang16}%
  \BibitemOpen
  \bibfield  {author} {\bibinfo {author} {\bibfnamefont {W.}~\bibnamefont
  {Kang}}, \bibinfo {author} {\bibfnamefont {Y.}~\bibnamefont {Huang}},
  \bibinfo {author} {\bibfnamefont {X.}~\bibnamefont {Zhang}}, \bibinfo
  {author} {\bibfnamefont {Y.}~\bibnamefont {Zhou}},\ and\ \bibinfo {author}
  {\bibfnamefont {W.}~\bibnamefont {Zhao}},\ }\bibfield  {title} {\bibinfo
  {title} {Skyrmion-electronics: An overview and outlook},\ }\href@noop {}
  {\bibfield  {journal} {\bibinfo  {journal} {Proc. IEEE}\ }\textbf {\bibinfo
  {volume} {104}},\ \bibinfo {pages} {2040} (\bibinfo {year}
  {2016})}\BibitemShut {NoStop}%
\bibitem [{\citenamefont {Fert}\ \emph {et~al.}(2013)\citenamefont {Fert},
  \citenamefont {Cros},\ and\ \citenamefont {Sampaio}}]{Fert13}%
  \BibitemOpen
  \bibfield  {author} {\bibinfo {author} {\bibfnamefont {A.}~\bibnamefont
  {Fert}}, \bibinfo {author} {\bibfnamefont {V.}~\bibnamefont {Cros}},\ and\
  \bibinfo {author} {\bibfnamefont {J.~a.}\ \bibnamefont {Sampaio}},\
  }\bibfield  {title} {\bibinfo {title} {Skyrmions on the track},\ }\href@noop
  {} {\bibfield  {journal} {\bibinfo  {journal} {Nat. Nanotechnol.}\ }\textbf
  {\bibinfo {volume} {8}},\ \bibinfo {pages} {152} (\bibinfo {year}
  {2013})}\BibitemShut {NoStop}%
\bibitem [{\citenamefont {M{\"u}hlbauer}\ \emph {et~al.}(2009)\citenamefont
  {M{\"u}hlbauer}, \citenamefont {Binz}, \citenamefont {Jonietz}, \citenamefont
  {Pfleiderer}, \citenamefont {Rosch}, \citenamefont {Neubauer}, \citenamefont
  {Georgii},\ and\ \citenamefont {B{\"o}ni}}]{Muhlbauer09}%
  \BibitemOpen
  \bibfield  {author} {\bibinfo {author} {\bibfnamefont {S.}~\bibnamefont
  {M{\"u}hlbauer}}, \bibinfo {author} {\bibfnamefont {B.}~\bibnamefont {Binz}},
  \bibinfo {author} {\bibfnamefont {F.}~\bibnamefont {Jonietz}}, \bibinfo
  {author} {\bibfnamefont {C.}~\bibnamefont {Pfleiderer}}, \bibinfo {author}
  {\bibfnamefont {A.}~\bibnamefont {Rosch}}, \bibinfo {author} {\bibfnamefont
  {A.}~\bibnamefont {Neubauer}}, \bibinfo {author} {\bibfnamefont
  {R.}~\bibnamefont {Georgii}},\ and\ \bibinfo {author} {\bibfnamefont
  {P.}~\bibnamefont {B{\"o}ni}},\ }\bibfield  {title} {\bibinfo {title}
  {Skyrmion lattice in a chiral magnet},\ }\href
  {https://doi.org/10.1126/science.1166767} {\bibfield  {journal} {\bibinfo
  {journal} {Science}\ }\textbf {\bibinfo {volume} {323}},\ \bibinfo {pages}
  {915} (\bibinfo {year} {2009})}\BibitemShut {NoStop}%
\bibitem [{\citenamefont {Li}\ \emph {et~al.}(2013)\citenamefont {Li},
  \citenamefont {Kanazawa}, \citenamefont {Yu}, \citenamefont {Tsukazaki},
  \citenamefont {Kawasaki}, \citenamefont {Ichikawa}, \citenamefont {Jin},
  \citenamefont {Kagawa},\ and\ \citenamefont {Tokura}}]{Li13}%
  \BibitemOpen
  \bibfield  {author} {\bibinfo {author} {\bibfnamefont {Y.}~\bibnamefont
  {Li}}, \bibinfo {author} {\bibfnamefont {N.}~\bibnamefont {Kanazawa}},
  \bibinfo {author} {\bibfnamefont {X.~Z.}\ \bibnamefont {Yu}}, \bibinfo
  {author} {\bibfnamefont {A.}~\bibnamefont {Tsukazaki}}, \bibinfo {author}
  {\bibfnamefont {M.}~\bibnamefont {Kawasaki}}, \bibinfo {author}
  {\bibfnamefont {M.}~\bibnamefont {Ichikawa}}, \bibinfo {author}
  {\bibfnamefont {X.~F.}\ \bibnamefont {Jin}}, \bibinfo {author} {\bibfnamefont
  {F.}~\bibnamefont {Kagawa}},\ and\ \bibinfo {author} {\bibfnamefont
  {Y.}~\bibnamefont {Tokura}},\ }\bibfield  {title} {\bibinfo {title} {Robust
  formation of skyrmions and topological hall effect anomaly in epitaxial thin
  films of \ce{ MnSi }},\ }\href
  {https://doi.org/10.1103/PhysRevLett.110.117202} {\bibfield  {journal}
  {\bibinfo  {journal} {Phys. Rev. Lett.}\ }\textbf {\bibinfo {volume} {110}},\
  \bibinfo {pages} {117202} (\bibinfo {year} {2013})}\BibitemShut {NoStop}%
\bibitem [{\citenamefont {Karube}\ \emph {et~al.}(2016)\citenamefont {Karube},
  \citenamefont {White}, \citenamefont {Reynolds}, \citenamefont {Gavilano},
  \citenamefont {Oike}, \citenamefont {Kikkawa}, \citenamefont {Kagawa},
  \citenamefont {Tokunaga}, \citenamefont {Rønnow}, \citenamefont {Tokura},\
  and\ \citenamefont {Taguchi}}]{Karube16}%
  \BibitemOpen
  \bibfield  {author} {\bibinfo {author} {\bibfnamefont {K.}~\bibnamefont
  {Karube}}, \bibinfo {author} {\bibfnamefont {J.~S.}\ \bibnamefont {White}},
  \bibinfo {author} {\bibfnamefont {N.}~\bibnamefont {Reynolds}}, \bibinfo
  {author} {\bibfnamefont {J.~L.}\ \bibnamefont {Gavilano}}, \bibinfo {author}
  {\bibfnamefont {H.}~\bibnamefont {Oike}}, \bibinfo {author} {\bibfnamefont
  {A.}~\bibnamefont {Kikkawa}}, \bibinfo {author} {\bibfnamefont
  {F.}~\bibnamefont {Kagawa}}, \bibinfo {author} {\bibfnamefont
  {Y.}~\bibnamefont {Tokunaga}}, \bibinfo {author} {\bibfnamefont {H.~M.}\
  \bibnamefont {Rønnow}}, \bibinfo {author} {\bibfnamefont {Y.}~\bibnamefont
  {Tokura}},\ and\ \bibinfo {author} {\bibfnamefont {Y.}~\bibnamefont
  {Taguchi}},\ }\bibfield  {title} {\bibinfo {title} {Robust metastable
  skyrmions and their triangular–square lattice structural transition in a
  high-temperature chiral magnet},\ }\href@noop {} {\bibfield  {journal}
  {\bibinfo  {journal} {Nat. Mater.}\ }\textbf {\bibinfo {volume} {15}},\
  \bibinfo {pages} {1237} (\bibinfo {year} {2016})}\BibitemShut {NoStop}%
\bibitem [{\citenamefont {Tokunaga}\ \emph {et~al.}(2015)\citenamefont
  {Tokunaga}, \citenamefont {Yu}, \citenamefont {R{\o}nnow}, \citenamefont
  {Morikawa}, \citenamefont {Taguchi},\ and\ \citenamefont
  {Tokura}}]{Tokunaga15}%
  \BibitemOpen
  \bibfield  {author} {\bibinfo {author} {\bibfnamefont {Y.}~\bibnamefont
  {Tokunaga}}, \bibinfo {author} {\bibfnamefont {J.~S.}\ \bibnamefont {Yu},
  \bibfnamefont {X.~Z.~White}}, \bibinfo {author} {\bibfnamefont {H.~M.}\
  \bibnamefont {R{\o}nnow}}, \bibinfo {author} {\bibfnamefont {D.}~\bibnamefont
  {Morikawa}}, \bibinfo {author} {\bibfnamefont {Y.}~\bibnamefont {Taguchi}},\
  and\ \bibinfo {author} {\bibfnamefont {Y.}~\bibnamefont {Tokura}},\
  }\bibfield  {title} {\bibinfo {title} {A new class of chiral materials
  hosting magnetic skyrmions beyond room temperature},\ }\href@noop {}
  {\bibfield  {journal} {\bibinfo  {journal} {Nat. Commun.}\ }\textbf {\bibinfo
  {volume} {6}},\ \bibinfo {pages} {7638} (\bibinfo {year} {2015})}\BibitemShut
  {NoStop}%
\bibitem [{\citenamefont {Peng}\ \emph {et~al.}(2018)\citenamefont {Peng},
  \citenamefont {Zhang}, \citenamefont {Ke}, \citenamefont {Kim}, \citenamefont
  {Zheng}, \citenamefont {Yan}, \citenamefont {Zhang}, \citenamefont {Gao},
  \citenamefont {Wang}, \citenamefont {Cai}, \citenamefont {Shen},
  \citenamefont {McQueeney}, \citenamefont {Kaminski}, \citenamefont {Kramer},\
  and\ \citenamefont {Zhou}}]{Peng18}%
  \BibitemOpen
  \bibfield  {author} {\bibinfo {author} {\bibfnamefont {L.}~\bibnamefont
  {Peng}}, \bibinfo {author} {\bibfnamefont {Y.}~\bibnamefont {Zhang}},
  \bibinfo {author} {\bibfnamefont {L.}~\bibnamefont {Ke}}, \bibinfo {author}
  {\bibfnamefont {T.-H.}\ \bibnamefont {Kim}}, \bibinfo {author} {\bibfnamefont
  {Q.}~\bibnamefont {Zheng}}, \bibinfo {author} {\bibfnamefont
  {J.}~\bibnamefont {Yan}}, \bibinfo {author} {\bibfnamefont {X.-G.}\
  \bibnamefont {Zhang}}, \bibinfo {author} {\bibfnamefont {Y.}~\bibnamefont
  {Gao}}, \bibinfo {author} {\bibfnamefont {S.}~\bibnamefont {Wang}}, \bibinfo
  {author} {\bibfnamefont {J.}~\bibnamefont {Cai}}, \bibinfo {author}
  {\bibfnamefont {B.}~\bibnamefont {Shen}}, \bibinfo {author} {\bibfnamefont
  {R.~J.}\ \bibnamefont {McQueeney}}, \bibinfo {author} {\bibfnamefont
  {A.}~\bibnamefont {Kaminski}}, \bibinfo {author} {\bibfnamefont {M.~J.}\
  \bibnamefont {Kramer}},\ and\ \bibinfo {author} {\bibfnamefont
  {L.}~\bibnamefont {Zhou}},\ }\bibfield  {title} {\bibinfo {title} {Relaxation
  dynamics of zero-field skyrmions over a wide temperature range},\ }\href
  {https://doi.org/10.1021/acs.nanolett.8b03553} {\bibfield  {journal}
  {\bibinfo  {journal} {Nano Letters}\ }\textbf {\bibinfo {volume} {18}},\
  \bibinfo {pages} {7777} (\bibinfo {year} {2018})}\BibitemShut {NoStop}%
\bibitem [{\citenamefont {Yu}\ \emph {et~al.}(2010)\citenamefont {Yu},
  \citenamefont {Kanazawa}, \citenamefont {Onose}, \citenamefont {Kimoto},
  \citenamefont {Zhang}, \citenamefont {Ishiwata}, \citenamefont {Matsui},\
  and\ \citenamefont {Tokura}}]{Yu10}%
  \BibitemOpen
  \bibfield  {author} {\bibinfo {author} {\bibfnamefont {X.~Z.}\ \bibnamefont
  {Yu}}, \bibinfo {author} {\bibfnamefont {N.}~\bibnamefont {Kanazawa}},
  \bibinfo {author} {\bibfnamefont {Y.}~\bibnamefont {Onose}}, \bibinfo
  {author} {\bibfnamefont {K.}~\bibnamefont {Kimoto}}, \bibinfo {author}
  {\bibfnamefont {W.~Z.}\ \bibnamefont {Zhang}}, \bibinfo {author}
  {\bibfnamefont {S.}~\bibnamefont {Ishiwata}}, \bibinfo {author}
  {\bibfnamefont {Y.}~\bibnamefont {Matsui}},\ and\ \bibinfo {author}
  {\bibfnamefont {Y.}~\bibnamefont {Tokura}},\ }\bibfield  {title} {\bibinfo
  {title} {Near room-temperature formation of a skyrmion crystal in thin-films
  of the helimagnet \ce{ FeGe }},\ }\href@noop {} {\bibfield  {journal}
  {\bibinfo  {journal} {Nat. Mater.}\ }\textbf {\bibinfo {volume} {10}},\
  \bibinfo {pages} {106} (\bibinfo {year} {2010})}\BibitemShut {NoStop}%
\bibitem [{\citenamefont {Li}\ \emph {et~al.}(2019)\citenamefont {Li},
  \citenamefont {Bykova}, \citenamefont {Zhang}, \citenamefont {Yu},
  \citenamefont {Tomasello}, \citenamefont {Carpentieri}, \citenamefont {Liu},
  \citenamefont {Guang}, \citenamefont {Gräfe}, \citenamefont {Weigand},
  \citenamefont {Burn}, \citenamefont {van~der Laan}, \citenamefont {Hesjedal},
  \citenamefont {Yan}, \citenamefont {Feng}, \citenamefont {Wan}, \citenamefont
  {Wei}, \citenamefont {Wang}, \citenamefont {Zhang}, \citenamefont {Xu},
  \citenamefont {Guo}, \citenamefont {Wei}, \citenamefont {Finocchio},
  \citenamefont {Han},\ and\ \citenamefont {Schütz}}]{Guoqiang19}%
  \BibitemOpen
  \bibfield  {author} {\bibinfo {author} {\bibfnamefont {W.}~\bibnamefont
  {Li}}, \bibinfo {author} {\bibfnamefont {I.}~\bibnamefont {Bykova}}, \bibinfo
  {author} {\bibfnamefont {S.}~\bibnamefont {Zhang}}, \bibinfo {author}
  {\bibfnamefont {G.}~\bibnamefont {Yu}}, \bibinfo {author} {\bibfnamefont
  {R.}~\bibnamefont {Tomasello}}, \bibinfo {author} {\bibfnamefont
  {M.}~\bibnamefont {Carpentieri}}, \bibinfo {author} {\bibfnamefont
  {Y.}~\bibnamefont {Liu}}, \bibinfo {author} {\bibfnamefont {Y.}~\bibnamefont
  {Guang}}, \bibinfo {author} {\bibfnamefont {J.}~\bibnamefont {Gräfe}},
  \bibinfo {author} {\bibfnamefont {M.}~\bibnamefont {Weigand}}, \bibinfo
  {author} {\bibfnamefont {D.~M.}\ \bibnamefont {Burn}}, \bibinfo {author}
  {\bibfnamefont {G.}~\bibnamefont {van~der Laan}}, \bibinfo {author}
  {\bibfnamefont {T.}~\bibnamefont {Hesjedal}}, \bibinfo {author}
  {\bibfnamefont {Z.}~\bibnamefont {Yan}}, \bibinfo {author} {\bibfnamefont
  {J.}~\bibnamefont {Feng}}, \bibinfo {author} {\bibfnamefont {C.}~\bibnamefont
  {Wan}}, \bibinfo {author} {\bibfnamefont {J.}~\bibnamefont {Wei}}, \bibinfo
  {author} {\bibfnamefont {X.}~\bibnamefont {Wang}}, \bibinfo {author}
  {\bibfnamefont {X.}~\bibnamefont {Zhang}}, \bibinfo {author} {\bibfnamefont
  {H.}~\bibnamefont {Xu}}, \bibinfo {author} {\bibfnamefont {C.}~\bibnamefont
  {Guo}}, \bibinfo {author} {\bibfnamefont {H.}~\bibnamefont {Wei}}, \bibinfo
  {author} {\bibfnamefont {G.}~\bibnamefont {Finocchio}}, \bibinfo {author}
  {\bibfnamefont {X.}~\bibnamefont {Han}},\ and\ \bibinfo {author}
  {\bibfnamefont {G.}~\bibnamefont {Schütz}},\ }\bibfield  {title} {\bibinfo
  {title} {Anatomy of skyrmionic textures in magnetic multilayers},\ }\href
  {https://doi.org/10.1002/adma.201807683} {\bibfield  {journal} {\bibinfo
  {journal} {Advanced Materials}\ }\textbf {\bibinfo {volume} {31}},\ \bibinfo
  {pages} {1807683} (\bibinfo {year} {2019})}\BibitemShut {NoStop}%
\bibitem [{\citenamefont {Soumyanarayanan}\ \emph {et~al.}(2017)\citenamefont
  {Soumyanarayanan}, \citenamefont {Raju}, \citenamefont {Gonzalez~Oyarce},
  \citenamefont {Tan}, \citenamefont {Im}, \citenamefont {Petrovi\'{c}},
  \citenamefont {Ho}, \citenamefont {Khoo}, \citenamefont {Tran}, \citenamefont
  {Gan}, \citenamefont {Ernult},\ and\ \citenamefont {Panagopoulos}}]{Anjan17}%
  \BibitemOpen
  \bibfield  {author} {\bibinfo {author} {\bibfnamefont {A.}~\bibnamefont
  {Soumyanarayanan}}, \bibinfo {author} {\bibfnamefont {M.}~\bibnamefont
  {Raju}}, \bibinfo {author} {\bibfnamefont {A.~L.}\ \bibnamefont
  {Gonzalez~Oyarce}}, \bibinfo {author} {\bibfnamefont {A.~K.~C.}\ \bibnamefont
  {Tan}}, \bibinfo {author} {\bibfnamefont {M.-Y.}\ \bibnamefont {Im}},
  \bibinfo {author} {\bibfnamefont {A.}~\bibnamefont {Petrovi\'{c}}}, \bibinfo
  {author} {\bibfnamefont {P.}~\bibnamefont {Ho}}, \bibinfo {author}
  {\bibfnamefont {K.~H.}\ \bibnamefont {Khoo}}, \bibinfo {author}
  {\bibfnamefont {M.}~\bibnamefont {Tran}}, \bibinfo {author} {\bibfnamefont
  {C.~K.}\ \bibnamefont {Gan}}, \bibinfo {author} {\bibfnamefont
  {F.}~\bibnamefont {Ernult}},\ and\ \bibinfo {author} {\bibfnamefont
  {C.}~\bibnamefont {Panagopoulos}},\ }\bibfield  {title} {\bibinfo {title}
  {Tunable room-temperature magnetic skyrmions in \ce{ Ir/Fe/Co/Pt }
  multilayers},\ }\href@noop {} {\bibfield  {journal} {\bibinfo  {journal}
  {Nat. Mater.}\ }\textbf {\bibinfo {volume} {16}},\ \bibinfo {pages} {898}
  (\bibinfo {year} {2017})}\BibitemShut {NoStop}%
\bibitem [{\citenamefont {Jiang}\ \emph {et~al.}(2015)\citenamefont {Jiang},
  \citenamefont {Upadhyaya}, \citenamefont {Zhang}, \citenamefont {Yu},
  \citenamefont {Jungfleisch}, \citenamefont {Fradin}, \citenamefont {Pearson},
  \citenamefont {Tserkovnyak}, \citenamefont {Wang}, \citenamefont {Heinonen},
  \citenamefont {te~Velthuis},\ and\ \citenamefont {Hoffmann}}]{Jiang15}%
  \BibitemOpen
  \bibfield  {author} {\bibinfo {author} {\bibfnamefont {W.}~\bibnamefont
  {Jiang}}, \bibinfo {author} {\bibfnamefont {P.}~\bibnamefont {Upadhyaya}},
  \bibinfo {author} {\bibfnamefont {W.}~\bibnamefont {Zhang}}, \bibinfo
  {author} {\bibfnamefont {G.}~\bibnamefont {Yu}}, \bibinfo {author}
  {\bibfnamefont {M.~B.}\ \bibnamefont {Jungfleisch}}, \bibinfo {author}
  {\bibfnamefont {F.~Y.}\ \bibnamefont {Fradin}}, \bibinfo {author}
  {\bibfnamefont {J.~E.}\ \bibnamefont {Pearson}}, \bibinfo {author}
  {\bibfnamefont {Y.}~\bibnamefont {Tserkovnyak}}, \bibinfo {author}
  {\bibfnamefont {K.~L.}\ \bibnamefont {Wang}}, \bibinfo {author}
  {\bibfnamefont {O.}~\bibnamefont {Heinonen}}, \bibinfo {author}
  {\bibfnamefont {S.~G.~E.}\ \bibnamefont {te~Velthuis}},\ and\ \bibinfo
  {author} {\bibfnamefont {A.}~\bibnamefont {Hoffmann}},\ }\bibfield  {title}
  {\bibinfo {title} {Blowing magnetic skyrmion bubbles},\ }\href
  {https://doi.org/10.1126/science.aaa1442} {\bibfield  {journal} {\bibinfo
  {journal} {Science}\ }\textbf {\bibinfo {volume} {349}},\ \bibinfo {pages}
  {283} (\bibinfo {year} {2015})}\BibitemShut {NoStop}%
\bibitem [{\citenamefont {Shibata}\ \emph {et~al.}(2015)\citenamefont
  {Shibata}, \citenamefont {Iwasaki}, \citenamefont {Kanazawa}, \citenamefont
  {Aizawa}, \citenamefont {Tanigaki}, \citenamefont {Shirai}, \citenamefont
  {Nakajima}, \citenamefont {Kubota}, \citenamefont {Kawasaki}, \citenamefont
  {Park}, \citenamefont {Shindo}, \citenamefont {Nagaosa},\ and\ \citenamefont
  {Tokura}}]{Shibata15}%
  \BibitemOpen
  \bibfield  {author} {\bibinfo {author} {\bibfnamefont {K.}~\bibnamefont
  {Shibata}}, \bibinfo {author} {\bibfnamefont {J.}~\bibnamefont {Iwasaki}},
  \bibinfo {author} {\bibfnamefont {N.}~\bibnamefont {Kanazawa}}, \bibinfo
  {author} {\bibfnamefont {S.}~\bibnamefont {Aizawa}}, \bibinfo {author}
  {\bibfnamefont {T.}~\bibnamefont {Tanigaki}}, \bibinfo {author}
  {\bibfnamefont {M.}~\bibnamefont {Shirai}}, \bibinfo {author} {\bibfnamefont
  {T.}~\bibnamefont {Nakajima}}, \bibinfo {author} {\bibfnamefont
  {M.}~\bibnamefont {Kubota}}, \bibinfo {author} {\bibfnamefont
  {M.}~\bibnamefont {Kawasaki}}, \bibinfo {author} {\bibfnamefont {H.~S.}\
  \bibnamefont {Park}}, \bibinfo {author} {\bibfnamefont {D.}~\bibnamefont
  {Shindo}}, \bibinfo {author} {\bibfnamefont {N.}~\bibnamefont {Nagaosa}},\
  and\ \bibinfo {author} {\bibfnamefont {Y.}~\bibnamefont {Tokura}},\
  }\bibfield  {title} {\bibinfo {title} {Large anisotropic deformation of
  skyrmions in strained crystal},\ }\href@noop {} {\bibfield  {journal}
  {\bibinfo  {journal} {Nat. Nanotechnol.}\ }\textbf {\bibinfo {volume} {10}},\
  \bibinfo {pages} {589} (\bibinfo {year} {2015})}\BibitemShut {NoStop}%
\bibitem [{\citenamefont {Chacon}\ \emph {et~al.}(2015)\citenamefont {Chacon},
  \citenamefont {Bauer}, \citenamefont {Adams}, \citenamefont {Rucker},
  \citenamefont {Brandl}, \citenamefont {Georgii}, \citenamefont {Garst},\ and\
  \citenamefont {Pfleiderer}}]{Chacon15}%
  \BibitemOpen
  \bibfield  {author} {\bibinfo {author} {\bibfnamefont {A.}~\bibnamefont
  {Chacon}}, \bibinfo {author} {\bibfnamefont {A.}~\bibnamefont {Bauer}},
  \bibinfo {author} {\bibfnamefont {T.}~\bibnamefont {Adams}}, \bibinfo
  {author} {\bibfnamefont {F.}~\bibnamefont {Rucker}}, \bibinfo {author}
  {\bibfnamefont {G.}~\bibnamefont {Brandl}}, \bibinfo {author} {\bibfnamefont
  {R.}~\bibnamefont {Georgii}}, \bibinfo {author} {\bibfnamefont
  {M.}~\bibnamefont {Garst}},\ and\ \bibinfo {author} {\bibfnamefont
  {C.}~\bibnamefont {Pfleiderer}},\ }\bibfield  {title} {\bibinfo {title}
  {Uniaxial pressure dependence of magnetic order in \ce{ MnSi }},\ }\href
  {https://doi.org/10.1103/PhysRevLett.115.267202} {\bibfield  {journal}
  {\bibinfo  {journal} {Phys. Rev. Lett.}\ }\textbf {\bibinfo {volume} {115}},\
  \bibinfo {pages} {267202} (\bibinfo {year} {2015})}\BibitemShut {NoStop}%
\bibitem [{\citenamefont {Nii}\ \emph {et~al.}(2015)\citenamefont {Nii},
  \citenamefont {Nakajima}, \citenamefont {Kikkawa}, \citenamefont {Yamasaki},
  \citenamefont {Ohishi}, \citenamefont {Suzuki}, \citenamefont {Taguchi},
  \citenamefont {Arima}, \citenamefont {Tokura},\ and\ \citenamefont
  {Iwasa}}]{Nii15}%
  \BibitemOpen
  \bibfield  {author} {\bibinfo {author} {\bibfnamefont {Y.}~\bibnamefont
  {Nii}}, \bibinfo {author} {\bibfnamefont {T.}~\bibnamefont {Nakajima}},
  \bibinfo {author} {\bibfnamefont {A.}~\bibnamefont {Kikkawa}}, \bibinfo
  {author} {\bibfnamefont {Y.}~\bibnamefont {Yamasaki}}, \bibinfo {author}
  {\bibfnamefont {K.}~\bibnamefont {Ohishi}}, \bibinfo {author} {\bibfnamefont
  {J.}~\bibnamefont {Suzuki}}, \bibinfo {author} {\bibfnamefont
  {Y.}~\bibnamefont {Taguchi}}, \bibinfo {author} {\bibfnamefont
  {T.}~\bibnamefont {Arima}}, \bibinfo {author} {\bibfnamefont
  {Y.}~\bibnamefont {Tokura}},\ and\ \bibinfo {author} {\bibfnamefont
  {Y.}~\bibnamefont {Iwasa}},\ }\bibfield  {title} {\bibinfo {title} {Uniaxial
  stress control of skyrmion phase},\ }\href@noop {} {\bibfield  {journal}
  {\bibinfo  {journal} {Nat. Commun.}\ }\textbf {\bibinfo {volume} {6}},\
  \bibinfo {pages} {8539} (\bibinfo {year} {2015})}\BibitemShut {NoStop}%
\bibitem [{\citenamefont {Seki}\ \emph {et~al.}(2017)\citenamefont {Seki},
  \citenamefont {Okamura}, \citenamefont {Shibata}, \citenamefont {Takagi},
  \citenamefont {Khanh}, \citenamefont {Kagawa}, \citenamefont {Arima},\ and\
  \citenamefont {Tokura}}]{Seki17}%
  \BibitemOpen
  \bibfield  {author} {\bibinfo {author} {\bibfnamefont {S.}~\bibnamefont
  {Seki}}, \bibinfo {author} {\bibfnamefont {Y.}~\bibnamefont {Okamura}},
  \bibinfo {author} {\bibfnamefont {K.}~\bibnamefont {Shibata}}, \bibinfo
  {author} {\bibfnamefont {R.}~\bibnamefont {Takagi}}, \bibinfo {author}
  {\bibfnamefont {N.~D.}\ \bibnamefont {Khanh}}, \bibinfo {author}
  {\bibfnamefont {F.}~\bibnamefont {Kagawa}}, \bibinfo {author} {\bibfnamefont
  {T.}~\bibnamefont {Arima}},\ and\ \bibinfo {author} {\bibfnamefont
  {Y.}~\bibnamefont {Tokura}},\ }\bibfield  {title} {\bibinfo {title}
  {Stabilization of magnetic skyrmions by uniaxial tensile strain},\ }\href
  {https://doi.org/10.1103/PhysRevB.96.220404} {\bibfield  {journal} {\bibinfo
  {journal} {Phys. Rev. B}\ }\textbf {\bibinfo {volume} {96}},\ \bibinfo
  {pages} {220404} (\bibinfo {year} {2017})}\BibitemShut {NoStop}%
\bibitem [{\citenamefont {Butenko}\ \emph {et~al.}(2010)\citenamefont
  {Butenko}, \citenamefont {Leonov}, \citenamefont {R\"o\ss{}ler},\ and\
  \citenamefont {Bogdanov}}]{Butenko10}%
  \BibitemOpen
  \bibfield  {author} {\bibinfo {author} {\bibfnamefont {A.~B.}\ \bibnamefont
  {Butenko}}, \bibinfo {author} {\bibfnamefont {A.~A.}\ \bibnamefont {Leonov}},
  \bibinfo {author} {\bibfnamefont {U.~K.}\ \bibnamefont {R\"o\ss{}ler}},\ and\
  \bibinfo {author} {\bibfnamefont {A.~N.}\ \bibnamefont {Bogdanov}},\
  }\bibfield  {title} {\bibinfo {title} {Stabilization of skyrmion textures by
  uniaxial distortions in noncentrosymmetric cubic helimagnets},\ }\href
  {https://doi.org/10.1103/PhysRevB.82.052403} {\bibfield  {journal} {\bibinfo
  {journal} {Phys. Rev. B}\ }\textbf {\bibinfo {volume} {82}},\ \bibinfo
  {pages} {052403} (\bibinfo {year} {2010})}\BibitemShut {NoStop}%
\bibitem [{\citenamefont {Wang}\ \emph {et~al.}(2018)\citenamefont {Wang},
  \citenamefont {Shi},\ and\ \citenamefont {Kamlah}}]{JWang18}%
  \BibitemOpen
  \bibfield  {author} {\bibinfo {author} {\bibfnamefont {J.}~\bibnamefont
  {Wang}}, \bibinfo {author} {\bibfnamefont {Y.}~\bibnamefont {Shi}},\ and\
  \bibinfo {author} {\bibfnamefont {M.}~\bibnamefont {Kamlah}},\ }\bibfield
  {title} {\bibinfo {title} {Uniaxial strain modulation of the skyrmion phase
  transition in ferromagnetic thin films},\ }\href
  {https://doi.org/10.1103/PhysRevB.97.024429} {\bibfield  {journal} {\bibinfo
  {journal} {Phys. Rev. B}\ }\textbf {\bibinfo {volume} {97}},\ \bibinfo
  {pages} {024429} (\bibinfo {year} {2018})}\BibitemShut {NoStop}%
\bibitem [{\citenamefont {Praetorius}\ \emph {et~al.}(2018)\citenamefont
  {Praetorius}, \citenamefont {Ruggeri},\ and\ \citenamefont
  {Stiftner}}]{Praetorius18}%
  \BibitemOpen
  \bibfield  {author} {\bibinfo {author} {\bibfnamefont {D.}~\bibnamefont
  {Praetorius}}, \bibinfo {author} {\bibfnamefont {M.}~\bibnamefont
  {Ruggeri}},\ and\ \bibinfo {author} {\bibfnamefont {B.}~\bibnamefont
  {Stiftner}},\ }\bibfield  {title} {\bibinfo {title} {Convergence of an
  implicit–explicit midpoint scheme for computational micromagnetics},\
  }\href {https://doi.org/https://doi.org/10.1016/j.camwa.2017.11.028}
  {\bibfield  {journal} {\bibinfo  {journal} {Computers \& Mathematics with
  Applications}\ }\textbf {\bibinfo {volume} {75}},\ \bibinfo {pages} {1719 }
  (\bibinfo {year} {2018})}\BibitemShut {NoStop}%
\bibitem [{\citenamefont {Alouges}\ \emph {et~al.}(2014)\citenamefont
  {Alouges}, \citenamefont {Kritsikis}, \citenamefont {Steiner},\ and\
  \citenamefont {Toussaint}}]{Alouges2014}%
  \BibitemOpen
  \bibfield  {author} {\bibinfo {author} {\bibfnamefont {F.}~\bibnamefont
  {Alouges}}, \bibinfo {author} {\bibfnamefont {E.}~\bibnamefont {Kritsikis}},
  \bibinfo {author} {\bibfnamefont {J.}~\bibnamefont {Steiner}},\ and\ \bibinfo
  {author} {\bibfnamefont {J.-C.}\ \bibnamefont {Toussaint}},\ }\bibfield
  {title} {\bibinfo {title} {A convergent and precise finite element scheme for
  {L}andau--{L}ifschitz--{G}ilbert equation},\ }\href
  {https://doi.org/10.1007/s00211-014-0615-3} {\bibfield  {journal} {\bibinfo
  {journal} {Numerische Mathematik}\ }\textbf {\bibinfo {volume} {128}},\
  \bibinfo {pages} {407} (\bibinfo {year} {2014})}\BibitemShut {NoStop}%
\bibitem [{\citenamefont {Gilbert}(2004)}]{Gilbert04}%
  \BibitemOpen
  \bibfield  {author} {\bibinfo {author} {\bibfnamefont {T.~L.}\ \bibnamefont
  {Gilbert}},\ }\bibfield  {title} {\bibinfo {title} {A phenomenological theory
  of damping in ferromagnetic materials},\ }\href@noop {} {\bibfield  {journal}
  {\bibinfo  {journal} {IEEE Trans. Magn.}\ }\textbf {\bibinfo {volume} {40}},\
  \bibinfo {pages} {3443} (\bibinfo {year} {2004})}\BibitemShut {NoStop}%
\bibitem [{\citenamefont {Eastman}(1966)}]{Eastman66}%
  \BibitemOpen
  \bibfield  {author} {\bibinfo {author} {\bibfnamefont {D.~E.}\ \bibnamefont
  {Eastman}},\ }\bibfield  {title} {\bibinfo {title} {Ultrasonic study of
  first-order and second-order magnetoelastic properties of yttrium iron
  garnet},\ }\href {https://doi.org/10.1103/PhysRev.148.530} {\bibfield
  {journal} {\bibinfo  {journal} {Phys. Rev.}\ }\textbf {\bibinfo {volume}
  {148}},\ \bibinfo {pages} {530} (\bibinfo {year} {1966})}\BibitemShut
  {NoStop}%
\bibitem [{\citenamefont {Zhang}\ and\ \citenamefont {Li}(2004)}]{ZhangLi04}%
  \BibitemOpen
  \bibfield  {author} {\bibinfo {author} {\bibfnamefont {S.}~\bibnamefont
  {Zhang}}\ and\ \bibinfo {author} {\bibfnamefont {Z.}~\bibnamefont {Li}},\
  }\bibfield  {title} {\bibinfo {title} {Roles of nonequilibrium conduction
  electrons on the magnetization dynamics of ferromagnets},\ }\href
  {https://doi.org/10.1103/PhysRevLett.93.127204} {\bibfield  {journal}
  {\bibinfo  {journal} {Phys. Rev. Lett.}\ }\textbf {\bibinfo {volume} {93}},\
  \bibinfo {pages} {127204} (\bibinfo {year} {2004})}\BibitemShut {NoStop}%
\bibitem [{\citenamefont {Garate}\ \emph {et~al.}(2009)\citenamefont {Garate},
  \citenamefont {Gilmore}, \citenamefont {Stiles},\ and\ \citenamefont
  {MacDonald}}]{Garate09}%
  \BibitemOpen
  \bibfield  {author} {\bibinfo {author} {\bibfnamefont {I.}~\bibnamefont
  {Garate}}, \bibinfo {author} {\bibfnamefont {K.}~\bibnamefont {Gilmore}},
  \bibinfo {author} {\bibfnamefont {M.~D.}\ \bibnamefont {Stiles}},\ and\
  \bibinfo {author} {\bibfnamefont {A.~H.}\ \bibnamefont {MacDonald}},\
  }\bibfield  {title} {\bibinfo {title} {Nonadiabatic spin-transfer torque in
  real materials},\ }\href {https://doi.org/10.1103/PhysRevB.79.104416}
  {\bibfield  {journal} {\bibinfo  {journal} {Phys. Rev. B}\ }\textbf {\bibinfo
  {volume} {79}},\ \bibinfo {pages} {104416} (\bibinfo {year}
  {2009})}\BibitemShut {NoStop}%
\bibitem [{\citenamefont {Nye}(2004)}]{Nye04}%
  \BibitemOpen
  \bibfield  {author} {\bibinfo {author} {\bibfnamefont {J.~F.}\ \bibnamefont
  {Nye}},\ }\href@noop {} {\emph {\bibinfo {title} {Physical Properties of
  Crystals: Their Representation by Tensors and Matrices}}}\ (\bibinfo
  {publisher} {Oxford University Press},\ \bibinfo {year} {2004})\ \bibinfo
  {note} {page 143}\BibitemShut {NoStop}%
\bibitem [{\citenamefont {Gercek}(2007)}]{Gercek07}%
  \BibitemOpen
  \bibfield  {author} {\bibinfo {author} {\bibfnamefont {H.}~\bibnamefont
  {Gercek}},\ }\bibfield  {title} {\bibinfo {title} {Poisson's ratio values for
  rocks},\ }\href
  {https://doi.org/https://doi.org/10.1016/j.ijrmms.2006.04.011} {\bibfield
  {journal} {\bibinfo  {journal} {International Journal of Rock Mechanics and
  Mining Sciences}\ }\textbf {\bibinfo {volume} {44}},\ \bibinfo {pages} {1 }
  (\bibinfo {year} {2007})}\BibitemShut {NoStop}%
\bibitem [{\citenamefont {Pfeiler}\ \emph {et~al.}(2020)\citenamefont
  {Pfeiler}, \citenamefont {Ruggeri}, \citenamefont {Stiftner}, \citenamefont
  {Exl}, \citenamefont {Hochsteger}, \citenamefont {Hrkac}, \citenamefont
  {Schöberl}, \citenamefont {Mauser},\ and\ \citenamefont
  {Praetorius}}]{Pfeiler18}%
  \BibitemOpen
  \bibfield  {author} {\bibinfo {author} {\bibfnamefont {C.-M.}\ \bibnamefont
  {Pfeiler}}, \bibinfo {author} {\bibfnamefont {M.}~\bibnamefont {Ruggeri}},
  \bibinfo {author} {\bibfnamefont {B.}~\bibnamefont {Stiftner}}, \bibinfo
  {author} {\bibfnamefont {L.}~\bibnamefont {Exl}}, \bibinfo {author}
  {\bibfnamefont {M.}~\bibnamefont {Hochsteger}}, \bibinfo {author}
  {\bibfnamefont {G.}~\bibnamefont {Hrkac}}, \bibinfo {author} {\bibfnamefont
  {J.}~\bibnamefont {Schöberl}}, \bibinfo {author} {\bibfnamefont {N.~J.}\
  \bibnamefont {Mauser}},\ and\ \bibinfo {author} {\bibfnamefont
  {D.}~\bibnamefont {Praetorius}},\ }\bibfield  {title} {\bibinfo {title}
  {Computational micromagnetics with commics},\ }\href
  {https://doi.org/https://doi.org/10.1016/j.cpc.2019.106965} {\bibfield
  {journal} {\bibinfo  {journal} {Computer Physics Communications}\ }\textbf
  {\bibinfo {volume} {248}},\ \bibinfo {pages} {106965} (\bibinfo {year}
  {2020})}\BibitemShut {NoStop}%
\bibitem [{\citenamefont {Di~Fratta}\ \emph {et~al.}(2019)\citenamefont
  {Di~Fratta}, \citenamefont {Pfeiler}, \citenamefont {Praetorius},
  \citenamefont {Ruggeri},\ and\ \citenamefont {Stiftner}}]{Fratta17}%
  \BibitemOpen
  \bibfield  {author} {\bibinfo {author} {\bibfnamefont {G.}~\bibnamefont
  {Di~Fratta}}, \bibinfo {author} {\bibfnamefont {C.-M.}\ \bibnamefont
  {Pfeiler}}, \bibinfo {author} {\bibfnamefont {D.}~\bibnamefont {Praetorius}},
  \bibinfo {author} {\bibfnamefont {M.}~\bibnamefont {Ruggeri}},\ and\ \bibinfo
  {author} {\bibfnamefont {B.}~\bibnamefont {Stiftner}},\ }\bibfield  {title}
  {\bibinfo {title} {{Linear second-order IMEX-type integrator for the (eddy
  current) Landau–Lifshitz–Gilbert equation}},\ }\bibfield  {journal}
  {\bibinfo  {journal} {IMA Journal of Numerical Analysis}\ }\href
  {https://doi.org/10.1093/imanum/drz046} {10.1093/imanum/drz046} (\bibinfo
  {year} {2019}),\ \Eprint
  {https://arxiv.org/abs/https://academic.oup.com/imajna/advance-article-pdf/doi/10.1093/imanum/drz046/31148986/drz046.pdf}
  {https://academic.oup.com/imajna/advance-article-pdf/doi/10.1093/imanum/drz046/31148986/drz046.pdf}
  \BibitemShut {NoStop}%
\bibitem [{ngs(2019)}]{ngsolve}%
  \BibitemOpen
  \href@noop {} {\bibinfo {title} {Netgen/ngsolve finite element library,
  https://ngsolve.org}} (\bibinfo {year} {2019})\BibitemShut {NoStop}%
\bibitem [{\citenamefont {Karhu}\ \emph {et~al.}(2012)\citenamefont {Karhu},
  \citenamefont {R\"o\ss{}ler}, \citenamefont {Bogdanov}, \citenamefont
  {Kahwaji}, \citenamefont {Kirby}, \citenamefont {Fritzsche}, \citenamefont
  {Robertson}, \citenamefont {Majkrzak},\ and\ \citenamefont
  {Monchesky}}]{Karhu12}%
  \BibitemOpen
  \bibfield  {author} {\bibinfo {author} {\bibfnamefont {E.~A.}\ \bibnamefont
  {Karhu}}, \bibinfo {author} {\bibfnamefont {U.~K.}\ \bibnamefont
  {R\"o\ss{}ler}}, \bibinfo {author} {\bibfnamefont {A.~N.}\ \bibnamefont
  {Bogdanov}}, \bibinfo {author} {\bibfnamefont {S.}~\bibnamefont {Kahwaji}},
  \bibinfo {author} {\bibfnamefont {B.~J.}\ \bibnamefont {Kirby}}, \bibinfo
  {author} {\bibfnamefont {H.}~\bibnamefont {Fritzsche}}, \bibinfo {author}
  {\bibfnamefont {M.~D.}\ \bibnamefont {Robertson}}, \bibinfo {author}
  {\bibfnamefont {C.~F.}\ \bibnamefont {Majkrzak}},\ and\ \bibinfo {author}
  {\bibfnamefont {T.~L.}\ \bibnamefont {Monchesky}},\ }\bibfield  {title}
  {\bibinfo {title} {Chiral modulations and reorientation effects in mnsi thin
  films},\ }\href {https://doi.org/10.1103/PhysRevB.85.094429} {\bibfield
  {journal} {\bibinfo  {journal} {Phys. Rev. B}\ }\textbf {\bibinfo {volume}
  {85}},\ \bibinfo {pages} {094429} (\bibinfo {year} {2012})}\BibitemShut
  {NoStop}%
\bibitem [{\citenamefont {Neubauer}\ \emph {et~al.}(2009)\citenamefont
  {Neubauer}, \citenamefont {Pfleiderer}, \citenamefont {Binz}, \citenamefont
  {Rosch}, \citenamefont {Ritz}, \citenamefont {Niklowitz},\ and\ \citenamefont
  {B\"oni}}]{Neubauer09}%
  \BibitemOpen
  \bibfield  {author} {\bibinfo {author} {\bibfnamefont {A.}~\bibnamefont
  {Neubauer}}, \bibinfo {author} {\bibfnamefont {C.}~\bibnamefont
  {Pfleiderer}}, \bibinfo {author} {\bibfnamefont {B.}~\bibnamefont {Binz}},
  \bibinfo {author} {\bibfnamefont {A.}~\bibnamefont {Rosch}}, \bibinfo
  {author} {\bibfnamefont {R.}~\bibnamefont {Ritz}}, \bibinfo {author}
  {\bibfnamefont {P.~G.}\ \bibnamefont {Niklowitz}},\ and\ \bibinfo {author}
  {\bibfnamefont {P.}~\bibnamefont {B\"oni}},\ }\bibfield  {title} {\bibinfo
  {title} {Topological hall effect in the ${A}$ phase of \ce{ MnSi}},\ }\href
  {https://doi.org/10.1103/PhysRevLett.102.186602} {\bibfield  {journal}
  {\bibinfo  {journal} {Phys. Rev. Lett.}\ }\textbf {\bibinfo {volume} {102}},\
  \bibinfo {pages} {186602} (\bibinfo {year} {2009})}\BibitemShut {NoStop}%
\bibitem [{\citenamefont {Vansteenkiste}\ \emph {et~al.}(2014)\citenamefont
  {Vansteenkiste}, \citenamefont {Leliaert}, \citenamefont {Dvornik},
  \citenamefont {Helsen}, \citenamefont {Garcia-Sanchez},\ and\ \citenamefont
  {Van~Waeyenberge}}]{Vansteenkiste14}%
  \BibitemOpen
  \bibfield  {author} {\bibinfo {author} {\bibfnamefont {A.}~\bibnamefont
  {Vansteenkiste}}, \bibinfo {author} {\bibfnamefont {J.}~\bibnamefont
  {Leliaert}}, \bibinfo {author} {\bibfnamefont {M.}~\bibnamefont {Dvornik}},
  \bibinfo {author} {\bibfnamefont {M.}~\bibnamefont {Helsen}}, \bibinfo
  {author} {\bibfnamefont {F.}~\bibnamefont {Garcia-Sanchez}},\ and\ \bibinfo
  {author} {\bibfnamefont {B.}~\bibnamefont {Van~Waeyenberge}},\ }\bibfield
  {title} {\bibinfo {title} {The design and verification of mumax3},\ }\href
  {https://doi.org/10.1063/1.4899186} {\bibfield  {journal} {\bibinfo
  {journal} {AIP Advances}\ }\textbf {\bibinfo {volume} {4}},\ \bibinfo {pages}
  {107133} (\bibinfo {year} {2014})}\BibitemShut {NoStop}%
\bibitem [{\citenamefont {Takagi}\ \emph {et~al.}(2017)\citenamefont {Takagi},
  \citenamefont {Morikawa}, \citenamefont {Karube}, \citenamefont {Kanazawa},
  \citenamefont {Shibata}, \citenamefont {Tatara}, \citenamefont {Tokunaga},
  \citenamefont {Arima}, \citenamefont {Taguchi}, \citenamefont {Tokura},\ and\
  \citenamefont {Seki}}]{Takagi17}%
  \BibitemOpen
  \bibfield  {author} {\bibinfo {author} {\bibfnamefont {R.}~\bibnamefont
  {Takagi}}, \bibinfo {author} {\bibfnamefont {D.}~\bibnamefont {Morikawa}},
  \bibinfo {author} {\bibfnamefont {K.}~\bibnamefont {Karube}}, \bibinfo
  {author} {\bibfnamefont {N.}~\bibnamefont {Kanazawa}}, \bibinfo {author}
  {\bibfnamefont {K.}~\bibnamefont {Shibata}}, \bibinfo {author} {\bibfnamefont
  {G.}~\bibnamefont {Tatara}}, \bibinfo {author} {\bibfnamefont
  {Y.}~\bibnamefont {Tokunaga}}, \bibinfo {author} {\bibfnamefont
  {T.}~\bibnamefont {Arima}}, \bibinfo {author} {\bibfnamefont
  {Y.}~\bibnamefont {Taguchi}}, \bibinfo {author} {\bibfnamefont
  {Y.}~\bibnamefont {Tokura}},\ and\ \bibinfo {author} {\bibfnamefont
  {S.}~\bibnamefont {Seki}},\ }\bibfield  {title} {\bibinfo {title} {Spin-wave
  spectroscopy of the {D}zyaloshinskii-{M}oriya interaction in room-temperature
  chiral magnets hosting skyrmions},\ }\href
  {https://doi.org/10.1103/PhysRevB.95.220406} {\bibfield  {journal} {\bibinfo
  {journal} {Phys. Rev. B}\ }\textbf {\bibinfo {volume} {95}},\ \bibinfo
  {pages} {220406} (\bibinfo {year} {2017})}\BibitemShut {NoStop}%
\bibitem [{\citenamefont {Bocarsly}\ \emph {et~al.}(2019)\citenamefont
  {Bocarsly}, \citenamefont {Heikes}, \citenamefont {Brown}, \citenamefont
  {Wilson},\ and\ \citenamefont {Seshadri}}]{Bocarsly19}%
  \BibitemOpen
  \bibfield  {author} {\bibinfo {author} {\bibfnamefont {J.~D.}\ \bibnamefont
  {Bocarsly}}, \bibinfo {author} {\bibfnamefont {C.}~\bibnamefont {Heikes}},
  \bibinfo {author} {\bibfnamefont {C.~M.}\ \bibnamefont {Brown}}, \bibinfo
  {author} {\bibfnamefont {S.~D.}\ \bibnamefont {Wilson}},\ and\ \bibinfo
  {author} {\bibfnamefont {R.}~\bibnamefont {Seshadri}},\ }\bibfield  {title}
  {\bibinfo {title} {Deciphering structural and magnetic disorder in the chiral
  skyrmion host materials \ce{ Co_xZn_yMn_z } ($x+y+z=20$)},\ }\href
  {https://doi.org/10.1103/PhysRevMaterials.3.014402} {\bibfield  {journal}
  {\bibinfo  {journal} {Phys. Rev. Materials}\ }\textbf {\bibinfo {volume}
  {3}},\ \bibinfo {pages} {014402} (\bibinfo {year} {2019})}\BibitemShut
  {NoStop}%
\bibitem [{\citenamefont {Bl\"ochl}(1994)}]{Bloch94}%
  \BibitemOpen
  \bibfield  {author} {\bibinfo {author} {\bibfnamefont {P.~E.}\ \bibnamefont
  {Bl\"ochl}},\ }\bibfield  {title} {\bibinfo {title} {Projector augmented-wave
  method},\ }\href {https://doi.org/10.1103/PhysRevB.50.17953} {\bibfield
  {journal} {\bibinfo  {journal} {Phys. Rev. B}\ }\textbf {\bibinfo {volume}
  {50}},\ \bibinfo {pages} {17953} (\bibinfo {year} {1994})}\BibitemShut
  {NoStop}%
\bibitem [{\citenamefont {Kresse}\ and\ \citenamefont
  {Furthmüller}(1996)}]{KRESSE199615}%
  \BibitemOpen
  \bibfield  {author} {\bibinfo {author} {\bibfnamefont {G.}~\bibnamefont
  {Kresse}}\ and\ \bibinfo {author} {\bibfnamefont {J.}~\bibnamefont
  {Furthmüller}},\ }\bibfield  {title} {\bibinfo {title} {Efficiency of
  ab-initio total energy calculations for metals and semiconductors using a
  plane-wave basis set},\ }\href
  {https://doi.org/https://doi.org/10.1016/0927-0256(96)00008-0} {\bibfield
  {journal} {\bibinfo  {journal} {Computational Materials Science}\ }\textbf
  {\bibinfo {volume} {6}},\ \bibinfo {pages} {15 } (\bibinfo {year}
  {1996})}\BibitemShut {NoStop}%
\bibitem [{\citenamefont {Perdew}\ \emph {et~al.}(1996)\citenamefont {Perdew},
  \citenamefont {Burke},\ and\ \citenamefont {Ernzerhof}}]{Perdew96}%
  \BibitemOpen
  \bibfield  {author} {\bibinfo {author} {\bibfnamefont {J.~P.}\ \bibnamefont
  {Perdew}}, \bibinfo {author} {\bibfnamefont {K.}~\bibnamefont {Burke}},\ and\
  \bibinfo {author} {\bibfnamefont {M.}~\bibnamefont {Ernzerhof}},\ }\bibfield
  {title} {\bibinfo {title} {Generalized gradient approximation made simple},\
  }\href {https://doi.org/10.1103/PhysRevLett.77.3865} {\bibfield  {journal}
  {\bibinfo  {journal} {Phys. Rev. Lett.}\ }\textbf {\bibinfo {volume} {77}},\
  \bibinfo {pages} {3865} (\bibinfo {year} {1996})}\BibitemShut {NoStop}%
\bibitem [{\citenamefont {Xie}\ \emph {et~al.}(2013)\citenamefont {Xie},
  \citenamefont {Thimmaiah}, \citenamefont {Lamsal}, \citenamefont {Liu},
  \citenamefont {Heitmann}, \citenamefont {Quirinale}, \citenamefont {Goldman},
  \citenamefont {Pecharsky},\ and\ \citenamefont {Miller}}]{Xie13}%
  \BibitemOpen
  \bibfield  {author} {\bibinfo {author} {\bibfnamefont {W.}~\bibnamefont
  {Xie}}, \bibinfo {author} {\bibfnamefont {S.}~\bibnamefont {Thimmaiah}},
  \bibinfo {author} {\bibfnamefont {J.}~\bibnamefont {Lamsal}}, \bibinfo
  {author} {\bibfnamefont {J.}~\bibnamefont {Liu}}, \bibinfo {author}
  {\bibfnamefont {T.~W.}\ \bibnamefont {Heitmann}}, \bibinfo {author}
  {\bibfnamefont {D.}~\bibnamefont {Quirinale}}, \bibinfo {author}
  {\bibfnamefont {A.~I.}\ \bibnamefont {Goldman}}, \bibinfo {author}
  {\bibfnamefont {V.}~\bibnamefont {Pecharsky}},\ and\ \bibinfo {author}
  {\bibfnamefont {G.~J.}\ \bibnamefont {Miller}},\ }\bibfield  {title}
  {\bibinfo {title} {$\beta$-\ce{Mn}-type \ce { Co_{8+x}Zn_{12-x} } as a defect
  cubic laves phase: Site preferences, magnetism, and electronic structure},\
  }\href {https://doi.org/10.1021/ic4009653} {\bibfield  {journal} {\bibinfo
  {journal} {Inorganic Chemistry}\ }\textbf {\bibinfo {volume} {52}},\ \bibinfo
  {pages} {9399} (\bibinfo {year} {2013})}\BibitemShut {NoStop}%
\bibitem [{\citenamefont {Nakajima}\ \emph {et~al.}(2019)\citenamefont
  {Nakajima}, \citenamefont {Karube}, \citenamefont {Ishikawa}, \citenamefont
  {Yonemura}, \citenamefont {Reynolds}, \citenamefont {White}, \citenamefont
  {R\o{}nnow}, \citenamefont {Kikkawa}, \citenamefont {Tokunaga}, \citenamefont
  {Taguchi}, \citenamefont {Tokura},\ and\ \citenamefont {Arima}}]{Nakajima19}%
  \BibitemOpen
  \bibfield  {author} {\bibinfo {author} {\bibfnamefont {T.}~\bibnamefont
  {Nakajima}}, \bibinfo {author} {\bibfnamefont {K.}~\bibnamefont {Karube}},
  \bibinfo {author} {\bibfnamefont {Y.}~\bibnamefont {Ishikawa}}, \bibinfo
  {author} {\bibfnamefont {M.}~\bibnamefont {Yonemura}}, \bibinfo {author}
  {\bibfnamefont {N.}~\bibnamefont {Reynolds}}, \bibinfo {author}
  {\bibfnamefont {J.~S.}\ \bibnamefont {White}}, \bibinfo {author}
  {\bibfnamefont {H.~M.}\ \bibnamefont {R\o{}nnow}}, \bibinfo {author}
  {\bibfnamefont {A.}~\bibnamefont {Kikkawa}}, \bibinfo {author} {\bibfnamefont
  {Y.}~\bibnamefont {Tokunaga}}, \bibinfo {author} {\bibfnamefont
  {Y.}~\bibnamefont {Taguchi}}, \bibinfo {author} {\bibfnamefont
  {Y.}~\bibnamefont {Tokura}},\ and\ \bibinfo {author} {\bibfnamefont
  {T.}~\bibnamefont {Arima}},\ }\bibfield  {title} {\bibinfo {title}
  {Correlation between site occupancies and spin-glass transition in skyrmion
  host \ce{ Co_{10-x/2} Zn_{10-x/2} Mn_x}},\ }\href
  {https://doi.org/10.1103/PhysRevB.100.064407} {\bibfield  {journal} {\bibinfo
   {journal} {Phys. Rev. B}\ }\textbf {\bibinfo {volume} {100}},\ \bibinfo
  {pages} {064407} (\bibinfo {year} {2019})}\BibitemShut {NoStop}%
\bibitem [{\citenamefont {Hori}\ \emph {et~al.}(2007)\citenamefont {Hori},
  \citenamefont {Shiraish},\ and\ \citenamefont {Ishii}}]{Hori07}%
  \BibitemOpen
  \bibfield  {author} {\bibinfo {author} {\bibfnamefont {T.}~\bibnamefont
  {Hori}}, \bibinfo {author} {\bibfnamefont {H.}~\bibnamefont {Shiraish}},\
  and\ \bibinfo {author} {\bibfnamefont {Y.}~\bibnamefont {Ishii}},\ }\bibfield
   {title} {\bibinfo {title} {Magnetic properties of $\beta$-\ce{ MnCoZn }
  alloys},\ }\href {https://doi.org/https://doi.org/10.1016/j.jmmm.2006.10.582}
  {\bibfield  {journal} {\bibinfo  {journal} {Journal of Magnetism and Magnetic
  Materials}\ }\textbf {\bibinfo {volume} {310}},\ \bibinfo {pages} {1820 }
  (\bibinfo {year} {2007})}\BibitemShut {NoStop}%
\bibitem [{\citenamefont {Ishizuka}\ and\ \citenamefont
  {Allman}(2005)}]{ishizuka_allman_2005}%
  \BibitemOpen
  \bibfield  {author} {\bibinfo {author} {\bibfnamefont {K.}~\bibnamefont
  {Ishizuka}}\ and\ \bibinfo {author} {\bibfnamefont {B.}~\bibnamefont
  {Allman}},\ }\bibfield  {title} {\bibinfo {title} {Phase measurement in
  electron microscopy using the transport of intensity equation},\ }\href
  {https://doi.org/10.1017/S1551929500051592} {\bibfield  {journal} {\bibinfo
  {journal} {Microscopy Today}\ }\textbf {\bibinfo {volume} {13}},\ \bibinfo
  {pages} {22–25} (\bibinfo {year} {2005})}\BibitemShut {NoStop}%
\bibitem [{\citenamefont {Yu}\ \emph {et~al.}(2018)\citenamefont {Yu},
  \citenamefont {Koshibae}, \citenamefont {Tokunaga}, \citenamefont {Shibata},
  \citenamefont {Taguchi}, \citenamefont {Nagaosa},\ and\ \citenamefont
  {Tokura}}]{Yu2018}%
  \BibitemOpen
  \bibfield  {author} {\bibinfo {author} {\bibfnamefont {X.~Z.}\ \bibnamefont
  {Yu}}, \bibinfo {author} {\bibfnamefont {W.}~\bibnamefont {Koshibae}},
  \bibinfo {author} {\bibfnamefont {Y.}~\bibnamefont {Tokunaga}}, \bibinfo
  {author} {\bibfnamefont {K.}~\bibnamefont {Shibata}}, \bibinfo {author}
  {\bibfnamefont {Y.}~\bibnamefont {Taguchi}}, \bibinfo {author} {\bibfnamefont
  {N.}~\bibnamefont {Nagaosa}},\ and\ \bibinfo {author} {\bibfnamefont
  {Y.}~\bibnamefont {Tokura}},\ }\bibfield  {title} {\bibinfo {title}
  {Transformation between meron and skyrmion topological spin textures in a
  chiral magnet},\ }\href {https://doi.org/10.1038/s41586-018-0745-3}
  {\bibfield  {journal} {\bibinfo  {journal} {Nature}\ }\textbf {\bibinfo
  {volume} {564}},\ \bibinfo {pages} {95} (\bibinfo {year} {2018})}\BibitemShut
  {NoStop}%
\bibitem [{\citenamefont {Jonietz}\ \emph {et~al.}(2010)\citenamefont
  {Jonietz}, \citenamefont {M{\"u}hlbauer}, \citenamefont {Pfleiderer},
  \citenamefont {Neubauer}, \citenamefont {M{\"u}nzer}, \citenamefont {Bauer},
  \citenamefont {Adams}, \citenamefont {Georgii}, \citenamefont {B{\"o}ni},
  \citenamefont {Duine}, \citenamefont {Everschor}, \citenamefont {Garst},\
  and\ \citenamefont {Rosch}}]{Jonietz1648}%
  \BibitemOpen
  \bibfield  {author} {\bibinfo {author} {\bibfnamefont {F.}~\bibnamefont
  {Jonietz}}, \bibinfo {author} {\bibfnamefont {S.}~\bibnamefont
  {M{\"u}hlbauer}}, \bibinfo {author} {\bibfnamefont {C.}~\bibnamefont
  {Pfleiderer}}, \bibinfo {author} {\bibfnamefont {A.}~\bibnamefont
  {Neubauer}}, \bibinfo {author} {\bibfnamefont {W.}~\bibnamefont
  {M{\"u}nzer}}, \bibinfo {author} {\bibfnamefont {A.}~\bibnamefont {Bauer}},
  \bibinfo {author} {\bibfnamefont {T.}~\bibnamefont {Adams}}, \bibinfo
  {author} {\bibfnamefont {R.}~\bibnamefont {Georgii}}, \bibinfo {author}
  {\bibfnamefont {P.}~\bibnamefont {B{\"o}ni}}, \bibinfo {author}
  {\bibfnamefont {R.~A.}\ \bibnamefont {Duine}}, \bibinfo {author}
  {\bibfnamefont {K.}~\bibnamefont {Everschor}}, \bibinfo {author}
  {\bibfnamefont {M.}~\bibnamefont {Garst}},\ and\ \bibinfo {author}
  {\bibfnamefont {A.}~\bibnamefont {Rosch}},\ }\bibfield  {title} {\bibinfo
  {title} {Spin transfer torques in \ce{MnSi} at ultralow current densities},\
  }\href {https://doi.org/10.1126/science.1195709} {\bibfield  {journal}
  {\bibinfo  {journal} {Science}\ }\textbf {\bibinfo {volume} {330}},\ \bibinfo
  {pages} {1648} (\bibinfo {year} {2010})}\BibitemShut {NoStop}%
\bibitem [{\citenamefont {Jiang}\ \emph {et~al.}(2016)\citenamefont {Jiang},
  \citenamefont {Zhang}, \citenamefont {Yu}, \citenamefont {Zhang},
  \citenamefont {Wang}, \citenamefont {Benjamin~Jungfleisch}, \citenamefont
  {Pearson}, \citenamefont {Cheng}, \citenamefont {Heinonen}, \citenamefont
  {Wang}, \citenamefont {Zhou}, \citenamefont {Hoffmann},\ and\ \citenamefont
  {te~Velthuis}}]{Jiang2016Magnus}%
  \BibitemOpen
  \bibfield  {author} {\bibinfo {author} {\bibfnamefont {W.}~\bibnamefont
  {Jiang}}, \bibinfo {author} {\bibfnamefont {X.}~\bibnamefont {Zhang}},
  \bibinfo {author} {\bibfnamefont {G.}~\bibnamefont {Yu}}, \bibinfo {author}
  {\bibfnamefont {W.}~\bibnamefont {Zhang}}, \bibinfo {author} {\bibfnamefont
  {X.}~\bibnamefont {Wang}}, \bibinfo {author} {\bibfnamefont {M.}~\bibnamefont
  {Benjamin~Jungfleisch}}, \bibinfo {author} {\bibfnamefont {J.~E.}\
  \bibnamefont {Pearson}}, \bibinfo {author} {\bibfnamefont {X.}~\bibnamefont
  {Cheng}}, \bibinfo {author} {\bibfnamefont {O.}~\bibnamefont {Heinonen}},
  \bibinfo {author} {\bibfnamefont {K.~L.}\ \bibnamefont {Wang}}, \bibinfo
  {author} {\bibfnamefont {Y.}~\bibnamefont {Zhou}}, \bibinfo {author}
  {\bibfnamefont {A.}~\bibnamefont {Hoffmann}},\ and\ \bibinfo {author}
  {\bibfnamefont {S.~G.~E.}\ \bibnamefont {te~Velthuis}},\ }\bibfield  {title}
  {\bibinfo {title} {Direct observation of the skyrmion hall effect},\ }\href
  {https://doi.org/10.1038/nphys3883} {\bibfield  {journal} {\bibinfo
  {journal} {Nature Physics}\ }\textbf {\bibinfo {volume} {13}},\ \bibinfo
  {pages} {162} (\bibinfo {year} {2016})},\ \bibinfo {note}
  {article}\BibitemShut {NoStop}%
\bibitem [{\citenamefont {Woo}\ \emph {et~al.}(2016)\citenamefont {Woo},
  \citenamefont {Litzius}, \citenamefont {Kr\"{u}ger}, \citenamefont {Im},
  \citenamefont {Caretta}, \citenamefont {Richter}, \citenamefont {Mann},
  \citenamefont {Krone}, \citenamefont {Reeve}, \citenamefont {Weigand},
  \citenamefont {Agrawal}, \citenamefont {Lemesh}, \citenamefont {Mawass},
  \citenamefont {Fischer}, \citenamefont {Kl\"{a}ui},\ and\ \citenamefont
  {Beach}}]{Woo16}%
  \BibitemOpen
  \bibfield  {author} {\bibinfo {author} {\bibfnamefont {S.}~\bibnamefont
  {Woo}}, \bibinfo {author} {\bibfnamefont {K.}~\bibnamefont {Litzius}},
  \bibinfo {author} {\bibfnamefont {B.}~\bibnamefont {Kr\"{u}ger}}, \bibinfo
  {author} {\bibfnamefont {M.-Y.}\ \bibnamefont {Im}}, \bibinfo {author}
  {\bibfnamefont {L.}~\bibnamefont {Caretta}}, \bibinfo {author} {\bibfnamefont
  {K.}~\bibnamefont {Richter}}, \bibinfo {author} {\bibfnamefont
  {M.}~\bibnamefont {Mann}}, \bibinfo {author} {\bibfnamefont {A.}~\bibnamefont
  {Krone}}, \bibinfo {author} {\bibfnamefont {R.~M.}\ \bibnamefont {Reeve}},
  \bibinfo {author} {\bibfnamefont {M.}~\bibnamefont {Weigand}}, \bibinfo
  {author} {\bibfnamefont {P.}~\bibnamefont {Agrawal}}, \bibinfo {author}
  {\bibfnamefont {I.}~\bibnamefont {Lemesh}}, \bibinfo {author} {\bibfnamefont
  {M.-A.}\ \bibnamefont {Mawass}}, \bibinfo {author} {\bibfnamefont
  {P.}~\bibnamefont {Fischer}}, \bibinfo {author} {\bibfnamefont
  {M.}~\bibnamefont {Kl\"{a}ui}},\ and\ \bibinfo {author} {\bibfnamefont
  {G.~S.~D.}\ \bibnamefont {Beach}},\ }\bibfield  {title} {\bibinfo {title}
  {Observation of room-temperature magnetic skyrmions and their current-driven
  dynamics in ultrathin metallic ferromagnets},\ }\href@noop {} {\bibfield
  {journal} {\bibinfo  {journal} {Nat. Mater.}\ }\textbf {\bibinfo {volume}
  {15}},\ \bibinfo {pages} {501} (\bibinfo {year} {2016})}\BibitemShut
  {NoStop}%
\bibitem [{\citenamefont {Sampaio}\ \emph {et~al.}(2013)\citenamefont
  {Sampaio}, \citenamefont {Cros}, \citenamefont {Rohart}, \citenamefont
  {Thiaville},\ and\ \citenamefont {Fert}}]{Sampaio13}%
  \BibitemOpen
  \bibfield  {author} {\bibinfo {author} {\bibfnamefont {J.}~\bibnamefont
  {Sampaio}}, \bibinfo {author} {\bibfnamefont {V.}~\bibnamefont {Cros}},
  \bibinfo {author} {\bibfnamefont {S.}~\bibnamefont {Rohart}}, \bibinfo
  {author} {\bibfnamefont {A.}~\bibnamefont {Thiaville}},\ and\ \bibinfo
  {author} {\bibfnamefont {A.}~\bibnamefont {Fert}},\ }\bibfield  {title}
  {\bibinfo {title} {Nucleation, stability and current-induced motion of
  isolated magnetic skyrmions in nanostructures},\ }\href@noop {} {\bibfield
  {journal} {\bibinfo  {journal} {Nat. Nanotechnol.}\ }\textbf {\bibinfo
  {volume} {8}},\ \bibinfo {pages} {839} (\bibinfo {year} {2013})}\BibitemShut
  {NoStop}%
\bibitem [{\citenamefont {Nan}\ \emph {et~al.}(2019)\citenamefont {Nan},
  \citenamefont {Hu}, \citenamefont {Dai}, \citenamefont {Emori}, \citenamefont
  {Wang}, \citenamefont {Hu}, \citenamefont {Matyushov}, \citenamefont {Chen},\
  and\ \citenamefont {Sun}}]{Nan19}%
  \BibitemOpen
  \bibfield  {author} {\bibinfo {author} {\bibfnamefont {T.}~\bibnamefont
  {Nan}}, \bibinfo {author} {\bibfnamefont {J.-M.}\ \bibnamefont {Hu}},
  \bibinfo {author} {\bibfnamefont {M.}~\bibnamefont {Dai}}, \bibinfo {author}
  {\bibfnamefont {S.}~\bibnamefont {Emori}}, \bibinfo {author} {\bibfnamefont
  {X.}~\bibnamefont {Wang}}, \bibinfo {author} {\bibfnamefont {Z.}~\bibnamefont
  {Hu}}, \bibinfo {author} {\bibfnamefont {A.}~\bibnamefont {Matyushov}},
  \bibinfo {author} {\bibfnamefont {L.-Q.}\ \bibnamefont {Chen}},\ and\
  \bibinfo {author} {\bibfnamefont {N.}~\bibnamefont {Sun}},\ }\bibfield
  {title} {\bibinfo {title} {A strain-mediated magnetoelectric-spin-torque
  hybrid structure},\ }\href {https://doi.org/10.1002/adfm.201806371}
  {\bibfield  {journal} {\bibinfo  {journal} {Advanced Functional Materials}\
  }\textbf {\bibinfo {volume} {29}},\ \bibinfo {pages} {1806371} (\bibinfo
  {year} {2019})}\BibitemShut {NoStop}%
\bibitem [{\citenamefont {Hu}\ \emph {et~al.}(2015)\citenamefont {Hu},
  \citenamefont {Yang}, \citenamefont {Wang}, \citenamefont {Huang},
  \citenamefont {Zhang}, \citenamefont {Chen},\ and\ \citenamefont
  {Nan}}]{Hu15}%
  \BibitemOpen
  \bibfield  {author} {\bibinfo {author} {\bibfnamefont {J.-M.}\ \bibnamefont
  {Hu}}, \bibinfo {author} {\bibfnamefont {T.}~\bibnamefont {Yang}}, \bibinfo
  {author} {\bibfnamefont {J.}~\bibnamefont {Wang}}, \bibinfo {author}
  {\bibfnamefont {H.}~\bibnamefont {Huang}}, \bibinfo {author} {\bibfnamefont
  {J.}~\bibnamefont {Zhang}}, \bibinfo {author} {\bibfnamefont {L.-Q.}\
  \bibnamefont {Chen}},\ and\ \bibinfo {author} {\bibfnamefont {C.-W.}\
  \bibnamefont {Nan}},\ }\bibfield  {title} {\bibinfo {title} {Purely
  electric-field-driven perpendicular magnetization reversal},\ }\href
  {https://doi.org/10.1021/nl504108m} {\bibfield  {journal} {\bibinfo
  {journal} {Nano Letters}\ }\textbf {\bibinfo {volume} {15}},\ \bibinfo
  {pages} {616} (\bibinfo {year} {2015})}\BibitemShut {NoStop}%
\bibitem [{\citenamefont {Buzzi}\ \emph {et~al.}(2013)\citenamefont {Buzzi},
  \citenamefont {Chopdekar}, \citenamefont {Hockel}, \citenamefont {Bur},
  \citenamefont {Wu}, \citenamefont {Pilet}, \citenamefont {Warnicke},
  \citenamefont {Carman}, \citenamefont {Heyderman},\ and\ \citenamefont
  {Nolting}}]{Buzzi13}%
  \BibitemOpen
  \bibfield  {author} {\bibinfo {author} {\bibfnamefont {M.}~\bibnamefont
  {Buzzi}}, \bibinfo {author} {\bibfnamefont {R.~V.}\ \bibnamefont
  {Chopdekar}}, \bibinfo {author} {\bibfnamefont {J.~L.}\ \bibnamefont
  {Hockel}}, \bibinfo {author} {\bibfnamefont {A.}~\bibnamefont {Bur}},
  \bibinfo {author} {\bibfnamefont {T.}~\bibnamefont {Wu}}, \bibinfo {author}
  {\bibfnamefont {N.}~\bibnamefont {Pilet}}, \bibinfo {author} {\bibfnamefont
  {P.}~\bibnamefont {Warnicke}}, \bibinfo {author} {\bibfnamefont {G.~P.}\
  \bibnamefont {Carman}}, \bibinfo {author} {\bibfnamefont {L.~J.}\
  \bibnamefont {Heyderman}},\ and\ \bibinfo {author} {\bibfnamefont
  {F.}~\bibnamefont {Nolting}},\ }\bibfield  {title} {\bibinfo {title} {Single
  domain spin manipulation by electric fields in strain coupled artificial
  multiferroic nanostructures},\ }\href
  {https://doi.org/10.1103/PhysRevLett.111.027204} {\bibfield  {journal}
  {\bibinfo  {journal} {Phys. Rev. Lett.}\ }\textbf {\bibinfo {volume} {111}},\
  \bibinfo {pages} {027204} (\bibinfo {year} {2013})}\BibitemShut {NoStop}%
\end{thebibliography}%

\end{document}